\documentclass[a4paper,twocolumn,11pt,accepted=2024-03-08]{quantumarticle}
\pdfoutput=1

\usepackage[numbers,sort&compress]{natbib}

\usepackage[T1]{fontenc}
\usepackage{cancel}
\usepackage{color,graphicx}
\usepackage{amsmath}
\usepackage{amsbsy}
\usepackage{amsthm}
\usepackage{bbm}
\usepackage{soul}

\usepackage{bm}
\usepackage{epsfig}
\usepackage{graphicx}
\usepackage{subfigure}
\usepackage{dcolumn}
\usepackage{color,epstopdf}
\usepackage{amscd}
\usepackage{amsfonts}
\usepackage{amssymb}
\usepackage{mathrsfs}
\usepackage{verbatim}
\usepackage{amsmath}
\usepackage[nointegrals]{wasysym}
\usepackage[utf8]{inputenc}
\usepackage[T1]{fontenc}
\usepackage{mathtools}
\usepackage[mathscr]{euscript}
\usepackage{dsfont}
\usepackage{lipsum}
\usepackage{hyperref}
\usepackage{cleveref}
\usepackage{mathtools}
\usepackage{dsfont}
\usepackage{cleveref}
\usepackage{bbold}
\usepackage{wasysym}
\usepackage{pgfcore}
\usepackage{pxpgfmark}
\usepackage{amsthm}
\usepackage[]{cases}

\newtheorem{theorem}{Theorem}
\newtheorem{lemma}[theorem]{Lemma}
\newtheorem{corollary}{Corollary}[theorem]

\DeclareMathOperator{\erfc}{erfc}

\DeclareMathOperator{\Var}{Var}

\newcommand{\be}{\begin{equation}}
\newcommand{\ee}{\end{equation}}
\newcommand{\baln}{\begin{align}}
\newcommand{\ealn}{\end{align}}
\newcommand{\ben}{\begin{equation*}}
\newcommand{\een}{\end{equation*}}

\DeclareUnicodeCharacter{0301}{\'{e}}

\long\def\symbolfootnote[#1]#2{\begingroup%
\def\thefootnote{\fnsymbol{footnote}}\footnote[#1]{#2}\endgroup}

\newcommand{\ex}[1]{\mathrm{e}^{#1}}

\newcommand\expect{\mathbb{E}}

\newcommand{\one}{\mathds{1}}
\newcommand{\E}{\mathbb{E}}

\newcommand{\braket}[1]{\langle {#1} \rangle}

\newcommand{\Tr}{\mathrm{Tr}}

\newcommand\harpl[1]{\mathstrut\mkern2.5mu#1\mkern-11mu\raise1.5ex%
  \hbox{$\scriptscriptstyle\leftharpoonup$}}

\usepackage{graphicx}
\newcommand\smallO{
  \mathchoice
    {{\scriptstyle\mathcal{O}}}
    {{\scriptstyle\mathcal{O}}}
    {{\scriptscriptstyle\mathcal{O}}}
    {\scalebox{.7}{$\scriptscriptstyle\mathcal{O}$}}
  }

\begin{document}

\title{Sequential hypothesis testing for continuously-monitored quantum systems}

\author{Giulio Gasbarri}
\email{giulio.gasbarri@uab.cat}
\affiliation{F\'isica Te\`orica: Informaci\'o i Fen\`omens Qu\`antics, Department de F\'isica, Universitat Aut\`onoma de Barcelona, 08193 Bellaterra (Barcelona), Spain}
\author{Matias Bilkis}
\email{matias.bilkis@uab.cat}
\affiliation{F\'isica Te\`orica: Informaci\'o i Fen\`omens Qu\`antics, Department de F\'isica, Universitat Aut\`onoma de Barcelona, 08193 Bellaterra (Barcelona), Spain}
\affiliation{Computer Vision Center, Universitat Aut\`onoma de Barcelona, Spain}
\author{Elisabet Roda-Salichs}
\email{elisabet.roda@uab.cat}
\affiliation{F\'isica Te\`orica: Informaci\'o i Fen\`omens Qu\`antics, Department de F\'isica, Universitat Aut\`onoma de Barcelona, 08193 Bellaterra (Barcelona), Spain}
\author{John Calsamiglia}
\email{john.calsamiglia@uab.cat}
\affiliation{F\'isica Te\`orica: Informaci\'o i Fen\`omens Qu\`antics, Department de F\'isica, Universitat Aut\`onoma de Barcelona, 08193 Bellaterra (Barcelona), Spain}

\begin{abstract}

We consider a quantum system that is being continuously monitored, giving rise to a measurement signal. From such a stream of data, information needs to be inferred about the underlying system's dynamics. Here we focus on hypothesis testing problems and put forward the usage of sequential strategies where the signal is analyzed in real-time, allowing the experiment to be concluded as soon
as the underlying hypothesis can be identified with a certified prescribed success probability. We analyze the performance of sequential tests by studying the stopping-time behavior, showing a considerable advantage over currently-used strategies based on a fixed predetermined measurement time.
\end{abstract}

\maketitle

\section{Introduction}

Continuously monitored quantum systems such as optomechanical~\cite{Aspelmeyer2014Cavity,millen2020optomechanics} systems or atomic magnetometers~\cite{kitching2011atomic,budker2007optical} are considered one of the most promising platforms for building quantum-enhanced and ultraprecise sensors~\cite{li2021cavity, kumar2021cavity, barzanjeh2022optomechanics}.  Despite the remarkable progress in the development of these devices~\cite{kitching2006chip, kitching2011atomic, abbott2016observation,mitchell2020colloquium, wang2023beating}, and the substantial body of theoretical research~\cite{wiseman1993quantum, wiseman2009quantum,forstner2012sensitivity,Tsang2012Continuous, gammelmark2013bayesian, jacobs2014quantum, molmer_hypothesis_2015, Albarelli_2017,kiilerich_hypothesis_2018,ralph2018dynamical, Jimenez2018Signal,liu2020quantum,mitchell2020colloquium,Am2021,marchese2023optomechanics}, conducted in this field, there remains a vast area to explore concerning effective strategies for leveraging the gathered time-series data, particularly in the context of protocols that necessitate real-time data assessment.

Traditional statistical inference approaches rely on processing measurement data only after the experiment, or multiple repeated experiments, have been completed ~\cite{trees_detection_2001}.
However, in the context of continuously monitored sensors and for many real-life applications, it is highly pertinent to consider sequential strategies that process resources on the fly and make decisions based on the single stream of data accumulated so far~\cite{Tarta}.

The primary objective of this article is to shed light on the application of sequential analysis methodologies in continuously monitored quantum systems, with a specific focus on one of the most fundamental primitives in statistical inference: binary hypothesis testing. { This work, thus represents the sequential counterpart of the previous works on binary hypothesis testing in continuously monitored quantum system \cite{Tsang2012Continuous,molmer_hypothesis_2015} that were so far restricted to detection strategies of a fixed predetermined duration. } By leveraging the benefits of these sequential methods, we aim to overcome the limitations of traditional post-experiment analysis approaches and enable more efficient and accurate extraction of information from continuously monitored quantum systems.

In a broad sense, standard hypothesis testing aims to assess the minimum probability of error $\epsilon$ when identifying the true hypothesis after using a fixed number $n$ of samples. For independent and identically distributed (IID) samples the error scales exponentially with the number of observations, $\epsilon_{n}\doteq\mathrm{e}^{-n R}$~\footnote{In this work we give several asymptotic results.
We will adopt this notation from asymptotic analysis. Given to sequences $a_{n}$, $b_{n}$:  Equality to first order in the exponent: $a_{n}\doteq b_{n}$ for $\lim_{n\to \infty}\frac{1}{n}\log\tfrac{a_{n}}{b_{n}}=0$.
Asymptotic equivalence: $a_{n}\sim b_{n}$ for $\lim_{n\to \infty}\frac{a_{n}}{b_{n}}=1$;
$a_{n}=\smallO(b_{n})$ for $\lim_{n\to \infty}\frac{a_{n}}{b_{n}}=0$; $a_{n}=\mathcal{O}(b_{n})$ for $\exists$ constants $c,m$ s.t. $|a_{n}|<c |b_{n}|$ $\forall n\geq m$.
\label{footAs}}, where the error $R$ rate depends on the precise setting and the two probability distributions $(p_{0(1)})$ underlying each of the hypotheses. Instead,  in sequential hypothesis testing, data is sampled sequentially until the moment when the true hypothesis can be identified with a prescribed probability of error $\epsilon$.  Since different measurement records convey different information, the number of samples required $n$ to reach an accurate enough guess, the so-called stopping time, is itself a stochastic variable. Wald \cite{wald2004sequential} proved that, for IID sampling, the mean stopping time scales as $\expect_{0}[{n}]\sim -\tfrac{\log\epsilon}{D(p_{0}\|p_{1})}$ as $\epsilon\to 0$, where  $D(p_{0}\| p_{1})$ { is the relative entropy and it satisfies $D(p_{0}\|p_{1}) > R$.  This finding highlights that sequential strategies offer substantial resource savings in a classical IID scenario.} 

The problem of sequential hypothesis testing has been recently tackled in the quantum realm \cite{QuantumVargas2021, li_optimal_2022}, in a setting where copies of a quantum system (either in $\rho=\rho_{0}$ or in $\rho=\rho_{1}$)
are provided on demand. The ultimate quantum bound on the mean stopping time (or the mean number of sampled copies) has been shown to follow the ``quantized version’' of Wald's result: naively exchanging probability distributions $p_{0(1)}(x)$ for quantum states $\rho_{0(1)}$,  i.e. $\expect_{0}[n]\sim -\tfrac{\log\epsilon}{D(\rho_{0}||\rho_{1})}$.

Here,  we study a very different quantum setting where, instead of performing a sequence of measurements on an increasing number of copies, we perform a (continuous) sequence of measurements on the \emph{very same quantum system}; and the question is to discriminate between two possible internal dynamics of the monitored system. We envision a quantum sensor, in particular an optomechanical sensor, whose dynamics are affected by external factors such as the presence of external mass or other forces, and the task is to detect this presence by observing a single (possibly long) measured signal.

As previously mentioned, the characteristic trait of sequential problems is that the horizon of observations is not fixed in advance; it is a stochastic variable. Under quite mild assumptions we will give an expression for the mean stopping time and characterize the stopping-time distribution for a wide class of continuously monitored quantum systems, with special focus on those described by Gaussian quantum states and Gaussian measurement statistics.

This work is structured as follows. In Sec.\ref{sec:hypothesis_testing} we revisit the hypothesis testing scenario, and introduce both the deterministic (fixed horizon) strategy and the sequential probability ratio test (SPRT); which are illustrated in Sec.~\ref{ssec:gaussian_iid} by means of a simple Gaussian IID case. We then move to continuously-monitored quantum systems, described in Sec.~\ref{sec:QMON}, and introduce the Gaussian model under study in Sec.~\ref{ssec:QMONgauss}. Our results are presented in Sec.~\ref{sec:QMONtest}. We begin by analyzing the stochastic evolution of the relevant statistic for hypothesis testing, i.e. log-likelihood ratio (LLR), and in Sec.~\ref{ssec:theorems} our analytical results for general continuous-monitoring quantum systems are presented. We prove some general results on the properties of the SPRT stopping times, under some mild assumptions on the statistical properties of the LLR. The case of Gaussian quantum systems is studied in detail in Sec.~\ref{ssec:QMONGAUSStest}, first by providing analytical results for the first moments of the LLR distribution, and secondly by numerically studying its behavior in order to assess its advantage of sequential strategies against deterministic ones. Finally, an outlook is given in Sec.~\ref{sec:outlook}

\section{Hypothesis testing}\label{sec:hypothesis_testing}

Let us now introduce the basic framework of standard binary hypothesis testing.
Here one counts with a sequence of $n$ observations $\mathcal{Y}_{n}=(y_{1},\dots,y_{n})$ and the goal is to identify, with a minimum probability of error, which of two given hypotheses ---called the null hypothesis ($h_{0}$) and alternative ($h_{1}$) hypothesis--- is responsible for generating this string of data.
It is inherent in this formulation that
there exists a (known) model that gives the probabilities $P_{0}(E)$ and $P_{1}(E)$ for an event $E$ to occur under hypothesis $h_{0}$ or $h_{1}$, respectively.
An inference strategy is determined by a decision function $d:\mathcal{Y}_{\tau}\to \{0,1\}$ that assigns a guessed hypothesis to each of the possible strings $\mathcal{Y}_{n}$.

It is customary to assess the performance of such strategies by the so-called type-I error or false positive, $\alpha_{1}=P_{0}(d=1)$, occurring when $h_{0}$ is rejected despite holding true, and type II error or false negative $\alpha_{0}=P_{1}(d=0)$, occurring when $h_{0}$ is accepted while the data has been generated in accordance with $h_{1}$. Since there is a trade-off between these two types of error one cannot minimize them independently. For this reason, one defines a single
 figure of merit, i.e. an optimality criterion for the different strategies, based on two standard approaches:
\begin{itemize}
\item  \textit{Symmetric} hypothesis testing: follows a Bayesian approach, where  prior probabilities $\{\pi_{0},\pi_{1}=1-\pi_{0}\}$ are assigned to each hypothesis so that the total probability of error can be computed as
\begin{align}
\label{eq:perr}
&P_{\mathrm{err}}=P_{0}(d=1)\pi_{0}+P_{1}(d=0)\pi_{1} \\
&\geq \sum_{\mathcal{Y}_{n}}  \min\{P_{0}(d(\mathcal{Y}_{n})=0)\pi_{0},P_{1}(d(\mathcal{Y}_{n})=1)\pi_{1}\}\nonumber
\end{align}
where the lower bound is attained by deciding for the most likely hypothesis, i.e. $d(\mathcal{Y}_{n})=0$ if $p_{0}(\mathcal{Y}_{n})\pi_{0} \geq p_{1}(\mathcal{Y}_{n})\pi_{1}$ and $d(\mathcal{Y}_{n})=1$ otherwise.
 {We use the lowercase notation to denote the probability densities, and reserve uppercase to denote the probability of events}.
\item  \textit{Asymmetric} hypothesis testing, where one is rather interested in minimizing only one of the two types of error whilst maintaining the other smaller than a predefined constant.
\end{itemize}

A central quantity in binary hypothesis testing is the log-likelihood ratio (LLR)
\begin{equation}\label{eq:LLR}
\ell(\mathcal{Y}_{t}):=\log\frac{p_{1}(\mathcal{Y}_{n})}{p_{0}(\mathcal{Y}_{n})}.
\end{equation}

Indeed, the Neyman-Pearson theorem \cite{cover2006} singles out the LLR test as an optimal one in the following sense: For  $a>0$ define the decision function (called likelihood test)
\begin{align}
d_{a}(\mathcal{Y}_{n})&=0 \mbox{ if } \ell(\mathcal{Y}_{n})\geq a   \nonumber\\
d_{a}(\mathcal{Y}_{n})& =1 \mbox{ if } \ell(\mathcal{Y}_{n})< a
\end{align}
with error probabilities
\begin{align}\label{eq:typeError}
	\alpha_{0}&= P_{1}(\ell<a)=\int_{-\infty}^{a}P_{1}(\ell=x) dx \nonumber\\
	\alpha_{1}&= P_{0}(\ell \geq a)=\int_{a}^{\infty}P_{0}(\ell=x) dx.
\end{align}
Given any other decision function $d'$  with error probabilities $\alpha_{0}'$ and $\alpha_{1}'$, if  $\alpha'_k \leq\alpha_{k}$ then  $\alpha'_{k\oplus 1}\geq\alpha_{k\oplus 1}$. The likelihood test with a suitable choice of threshold $a$ is also optimal for symmetric and asymmetric hypothesis testing scenarios defined above. That is, the (log) likelihood function $\ell(\mathcal{Y}_{n})$ is the relevant statistic to define the acceptance (and rejection) region in the observation space.

Equipped with these definitions one can study the behavior of the probability of error $\epsilon_{n}$ as we increase the number of observations $n$. For IID samplings, the error can be seen to decay exponentially, and closed expressions can be found for the error rate  $R:=-\lim_{n\to\infty}\tfrac{1}{n}\log\epsilon_{n}$  in terms of the underlying PDF's  ---see the Chernoff bound and Stein lemma, e.g. in \cite{cover2006},  for the rates corresponding symmetric and asymmetric setting respectively.

Let us now move to \emph{sequential hypothesis testing}. Here a general inference strategy is defined as the duple $\mathcal{S}=\{\tau,d(\mathcal{Y}_{\tau})\}$ of a stopping time and a decision function. The stopping time shall be determined solely by the information available at each step of the process, or more succinctly: $\tau$ is a stopping time if, for any $n$, the occurrence of the event $\tau\leq n$ can be determined from $\mathcal{Y}_{n}$. The decision function takes as input the observed sequence until the stopping time $\tau$ and produces a guess for the hypothesis $d(\mathcal{Y}_{\tau})=\{0,1\}$.

Note that this set of strategies includes the traditional deterministic strategies, i.e. those where the stopping time is a fixed deterministic variable $\tau = n$,  where $n$ is a predetermined value. By relaxing this constraint and enabling samples to be provided on demand, we can explore new scenarios and utilize resources more efficiently. In contrast to the standard (deterministic) settings where given a number of observations,  one assesses the expected probability of error, we will assess the mean number of observations required in a sequential strategy to attain given error bounds.

We start by considering, the \emph{strong error conditions} \cite{Wald1945Sequential, slussarenko_quantum_2017,QuantumVargas2021}, a Bayesian scenario where both decisions ($d=0,1$) have a certified error below a given threshold, for each possible measurement record. We will see that this leads to the sequential probability ratio test (SPRT), a test that turns out to be optimal also in other settings (including non-Bayesian, asymmetric ones).

 If hypothesis $h_{k}$ is given with prior probability  $\pi_{k}$,
 after observing a measurement sequence $\mathcal{Y}_{t}$, the posterior probability is given by Bayes' update rule,
\begin{equation}\label{eq:str_cond}
p(h_{k}|\mathcal{Y}_{t}) = \frac{p(\mathcal{Y}_{t}|h_{k})\pi_{k}}{p(\mathcal{Y}_{t}|h_{0})\pi_{0}+p(\mathcal{Y}_{t}|h_{1})\pi_{1}}
\end{equation}
where $p(\mathcal{Y}_{t}|h_{k})=p_{k}(\mathcal{Y}_{t})$.
Hence, the strong error conditions mean that a decision can be taken only if either
\be\label{eq:strong_cond}
p(h_{0}|\mathcal{Y}_{t}) \geq 1-\epsilon_{0} \mbox{ or }
 p(h_{1}|\mathcal{Y}_{t}) \geq 1-\epsilon_{1}
\ee
hold \footnote{Note that the strong error conditions cannot be met for all possible outcomes of a deterministic test  {where the length of the sequence is finite and fixed in advance. Indeed, in a fixed-length scenario there is always a chance that the encountered sequence is not informative enough, i.e. its conditional probability does not   fulfill~\eqref{eq:strong_cond}.}}.
We will denote by $\mathbf{S}(\epsilon_{0},\epsilon_{1})$ the class of inference strategies $\mathcal{S}=\{d,\tau\}$ which satisfy prescribed strong error bounds \eqref{eq:strong_cond}.
The shortest permissible stopping time is the earliest moment at which either of these conditions is met. It is easy to see that the conditions in \eqref{eq:strong_cond} can be expressed in terms of the log-likelihood ratio \eqref{eq:LLR} as
\begin{align}\label{eq:link}
 \ell_{t}&\leq -a_{0}: = - \log\left(  \frac{1-\epsilon_{0}}{\epsilon_{0}} \frac{\pi_0}{\pi_1} \right),\nonumber\\
  \ell_{t}&\geq a_{1} :=  \log\left(\frac{1-\epsilon_{1}}{\epsilon_{1}}\frac{\pi_1}{\pi_0}\right).
  \end{align}
with $\ell_{t}:=\ell(\mathcal{Y}_{t})$. Therefore the optimal strategy (i.e. with the shortest stopping time) respecting the strong error conditions is given by an SPRT with the threshold values given in  \eqref{eq:link}. An SPRT is defined as follows,
by the stopping time
\begin{align}
 \tau_{\mathrm{s}} = \inf\{t \ge0 ; \ell_{n}\notin (-a_{0},a_{1})\}
\end{align}
and the decision function
\begin{align}
d_{\mathrm{s}}=\begin{cases}
1 &\textit{if}\quad \ell_{\tau_{\mathrm{s}}}\ge a_{1}\\
0 & \textit{if}\quad \ell_{\tau_{\mathrm{s}}}\le -a_{0}\\
\end{cases}
\end{align}
for some threshold values $a_{i}>0$. In plain words, a sequential probability test operates by following these rules
\begin{itemize}
  \item if $\ell(\mathcal{Y}_{t}) \ge a_{1}$, the test stops and  $h_{1}$ is accepted
  \item if $\ell(\mathcal{Y}_{t})\le - a_{0}$,  the test stops and $h_{0}$ is accepted
  \item if $a_{0} < \ell(\mathcal{Y}_{t}) < a_{1}$, the test continues by asking for new samples.
\end{itemize}
Clearly, if the strong error bounds  \eqref{eq:strong_cond} are the same for both hypothesis, $\epsilon_{0}=\epsilon_{1}=\epsilon$, then the unconditional total error probability
$P_{\mathrm{err}}$ \eqref{eq:perr}  is also bounded by $\epsilon$.
In the seminal works introducing the SPRT \cite{Wald1945Sequential,wald2004sequential}, Wald  showed that for an SPRT with thresholds $(-a_{0},a_{1})$ the type I and type II errors (that are defined with no need of Bayesian priors)  satisfy
\begin{align}\label{eq:pe_SPRT}
    \alpha_{1} &= P_{0}(\ell_{t} \ge a_{1})\le P_{1}(\ell \ge a_{1})\ex{-a_{1}} = (1-\alpha_{0})\ex{-a_{1}},\nonumber\\
    \alpha_{0} &=  P_{1}(\ell \le -a_{0}) \le P_{0}(\ell \le -a_{0}) \ex{-a_{0}} = (1-\alpha_{1})\ex{-a_{0}}.
\end{align}
In case of no (or negligible) overshooting, i.e. when the sampling process exactly stops when the decision contour is reached ($\ell_{\tau}=a_{0}$ or $\ell_{\tau}=a_{1}$), the inequalities are saturated and the probabilities of type I and II are given by
\begin{align}\label{eq:statistial_errors}
\alpha_{0} = \frac{1-\ex{-a_{1}}}{\ex{a_{0}}-\ex{-a_{1}}} \hspace{0.5cm}  \alpha_{1} = \frac{1-\ex{-a_{0}}}{\ex{a_{1}}-\ex{-a_{0}}}.
\end{align}

Throughout this work, we will denote by $\mathbf{C}(\alpha_{0},\alpha_{1})$  the class of inference strategies $\mathcal{S}$ which satisfy prescribed bounds on type I and type II error probabilities, $P_{0}(d=1)\leq \alpha_{1}$ and  $P_{1}(d=0)\leq \alpha_{0}$.
To distinguish those strategies, from those obeying the Bayesian strong error conditions (single-trajectory),  we will say that strategies in $\mathbf{C}(\alpha_{0},\alpha_{1})$ obey the \emph{weak error conditions}.
The Wald-Wolfowitz theorem \cite{wald_optimum_1948} establishes the optimality of the SPRT under IID,  in the sense that it minimizes the expected stopping time under both hypotheses, $\expect_{0}[{\tau}]$ and $\expect_{1}[{\tau}]$, among all tests in $\mathbf{C}(\alpha_{0},\alpha_{1})$ (sequential or not).
 In addition, for the IID setting the mean stopping time $\expect_{k}[{\tau}]$ can be computed, and can be seen to be significantly shorter than the time (number of samples) required for a deterministic bound to attain the same error bounds.

In the next section, we give an explicit example demonstrating the advantage of sequential hypothesis testing and showcasing some features that will also be present in continuously monitored quantum systems.

\subsection{Gaussian Distribution IID}\label{ssec:gaussian_iid}
Here, we offer a straightforward illustrative example in the IID case, where all computations can be performed analytically, which will also
serve to further introduce some essential results in hypothesis testing.

Let us assume that our observations are described by a random variable that, depending on the hypothesis, obeys
\begin{align}
h_{0}&: y= m_{0}+ \zeta, \nonumber\\
h_{1}&: y = m_{1}+\zeta \nonumber
\end{align}
where $\zeta$ is a Gaussian stochastic variable  with zero mean and variance $\mathbb{E}[\zeta^{2}]=\sigma^{2}$.

Since the probability density function (PDF) of the observations obeys $p_{k}(\mathcal{Y}_{t})=\prod_{i=1}^{t} p_{k}(y_{i})$ where $p_{k}(y_{i})$ are Gaussian for both hypothesis, $k=0,1$, it immediately follows  that the LLR is given by
\begin{align}
\label{eq:ellGauss}
\ell(\mathcal{Y}_{t}) &=\sum_{i=1}^{t}\ell(y_{i})= \frac{1}{2\sigma^{2}}\sum_{i=1}^{t} (m_{1}-y_{i})^{2} - (m_{0}-y_{i})^{2}\nonumber\\
&= \frac{({m}_{1}-{m}_{0})}{\sigma^{2}}  Y -  \frac{{m}_{1}^{2}- {m}_{0}^{2}}{ 2\sigma^{2}},
\end{align}
where  ${Y}_{t}= \sum_{i=1}^{t} y_{i}$, is Gaussian with mean $\expect_{k}[{Y}_{t}]=t\,m_{k}$ and variance $\Var_{k} Y_t=t\sigma^{2}$. That is, the LLR at a given time $t$, $\ell_{t}=\ell(\mathcal{Y}_{t})$, is itself a Gaussian random variable with mean and variance given by:
\begin{align}
\expect_{k}[\ell_{t}]&=: (-1)^{k\oplus 1}t\, \mu_{k}\; \mbox { with } \mu_{k}=\frac{(m_{1}-m_{0})^{2}}{2\sigma^{2}}\nonumber\\
\Var_{k}[\ell_{t}] &=: t\, \nu_{k}\; \mbox { with } \nu_{k}= \frac{({m}_{1}-{m}_{0})^{2}}{\sigma^{2}}
\end{align}
where, in order to present the results for both hypotheses $k\in\{0,1\}$ in a unified manner, here,  and throughout this paper, we utilize the modulo 2 addition so that $k\oplus 1$ represents the complementary hypothesis of $k$.
We note that both the mean and variance grow linearly with
the number of samples. More interestingly, note that all rates depend on a single parameter: $\mu_{1}=\mu_{0}=\mu= \frac{(m_{1}-m_{0})^{2}}{2\sigma^{2}}$ and $\nu_{0}=\nu_{1}=2 \mu$. This relation is not accidental, but it can be seen as a necessary condition for any setting with a Gaussian LLR (see Theorem \ref{th:gaussian_iid} below).

Having a full characterization of the distribution of the LLR statistic $\ell_{t}$, we can easily compute the type I and type II errors after a fixed number of samples $n$. From  \eqref{eq:typeError}  these errors can be obtained by integrating the tails of a Gaussian:
\begin{align}\label{eq:typeErrorG}
	\alpha_{0}&=
	\frac{1}{2} \erfc\left[\frac{\mu_{1} t-a}{\sqrt{2 \nu_{1} {t}}}\right]=\frac{1}{2} \erfc\left[\frac{{t} \mu -a}{2\sqrt{\mu {t}} }\right]\nonumber\\
	\alpha_{1}&=\frac{1}{2} \erfc\left[\frac{\mu_{0} {t}+a}{\sqrt{2{t}\nu_{0}} }\right]=\frac{1}{2} \erfc\left[\frac{{t} \mu +a}{2\sqrt{\mu {t}} }\right]\ .
\end{align}

\begin{figure}[t!]
\centering
\includegraphics[width=0.4\textwidth]{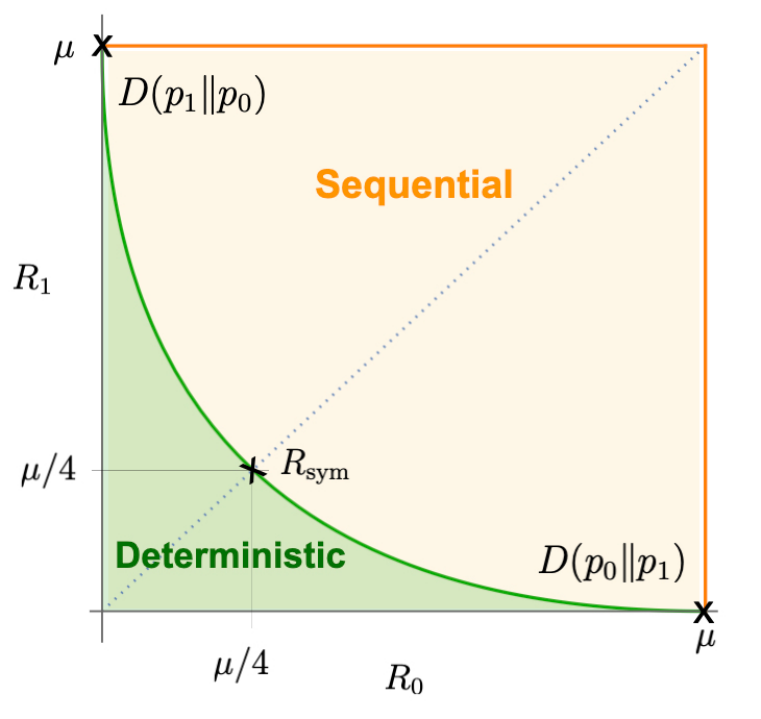}
\caption{\textit{Attainable asymptotic error rates (ratio between $\log \epsilon$ and
sampling time) for type I and type II error rates, for deterministic strategies (in green), and for sequential strategies (in orange).}}
\label{fig:ROC}
\end{figure}

By choosing a threshold value adapted to the number of samples available $a=\xi t$
(with $-\mu<\xi<\mu$) we get the asymptotic error rates
\be
\label{eq:deterr}
R_{k}^{(\xi)}:= \lim_{t\to \infty} -\frac{\log\alpha_{k}{(t)}}{t}
=\frac{(\mu+(-1)^{k}\xi)^{2}}{4\mu},
\ee
where we have used that for $b>0$, $\lim_{t\to \infty} \tfrac{1}{t}\log[\erfc(b \sqrt{t})]=-b^{2}$.
 Figure \ref{fig:ROC} shows, shaded in green, the attainable asymptotic error rates $(R_{0}, R_{1})$ for deterministic strategies. The optimal deterministic strategies are given by the curve $(R_{0}^{{(\xi)}}, R_{1}^{{(\xi)}})$ given by \eqref{eq:deterr}, which describes the optimal trade-off between the false positive and false negative error rates.

The point where both error rates coincide, $R_{0}^{{(\xi^{*})}}=R_{1}^{{(\xi^{*})}}=\frac{\mu}{4}$ corresponds to the symmetric (Bayesian) error rate,
\be
\label{eq:detersym}
R_{\mathrm{sym}}:= \lim_{t\to \infty} -\frac{\log P_{\mathrm{err}}{(t)}}{t}
=\frac{\mu}{4}
\ee
 i.e.
$P_{\mathrm{err}}\doteq \ex{-\mu t /4}$ ---independently of the choice of priors. Indeed,  from \eqref{eq:perr} it follows that the optimal threshold in this setting is $a=\log\tfrac{\pi_{0}}{\pi_{1}}$ for all $t$, which means that $\xi=a/t\to 0$.

For $\xi\to\pm\mu$ we get the optimal error rate for asymmetric hypothesis testing in agreement with Stein's lemma \cite{cover2006}: if $\alpha_{0}^{\epsilon}(t)=\min_d\{\alpha_{0}(t) \mbox{ such that } \alpha_{1}(t)<\epsilon\}$ for some fixed $\epsilon>0$, then the corresponding error rate is given by the relative entropy:
\begin{align}
R_{0}^{*}&:=-\lim _{t \rightarrow \infty} \frac{1}{t} \log \alpha_{0}^{\epsilon}(t)
\nonumber \\
&= {D\left(p_{0} \| p_{1}\right)}:=\int dy p_{0}(y)\log\tfrac{p_{0}(y)}{p_{1}(y)}=\nonumber\\
&=-\expect_{0}(\ell_{1})=\mu.
\label{eq:exl1}
\end{align}
Similarly, if we bound the type II error, $\alpha_{0}(t)<\epsilon$,  the optimal rate for the type I error will be  $R_{1}^{*}=D\left(P_{1} \| P_{0}\right)$.
It is worth mentioning that the relative entropy gives the fastest error decay rate one can attain, in the sense that any strategy that detects hypothesis $h_{k}$ with a smaller error, i.e.  {$R_{k}>D(p_{k}\|p_{k\oplus 1})$},  will effectively always decide for $h_{k}$, i.e. $\alpha_{k\oplus 1}(t)\to 1$ as $t\to\infty$ ---as can be also seen from \eqref{eq:typeErrorG} by taking $|\xi|>\mu$).

Let us now compute the average number of resources needed to achieve the same error probability with the SPRT. In the IID case, one can relate the average stopping time to the average LLR (of a single symbol) by first writing
\begin{align}
\expect_{k}[\ell(\mathcal{Y}_{\tau})] =\expect_{k}[
\sum_{i=1}^{\tau}\ell(y_{i})]=\expect_{k}[\ell(y)] \expect_{k}[\tau],
\end{align}
where the Wald identity \cite{wald2004sequential} is used to compute the expected value of a summation, where the range is determined by a stopping time (random variable).

In addition, if we neglect the overshooting, we can assume that at the stopping time,
the LLR will take one of the two possible values at the decision boundary, $\ell_{\tau}=(-1)^{k\oplus 1 }a_{k}$. Under hypothesis $h_{k}$ hitting the boundary $(-1)^{k} a_{k\oplus 1}$ is associated to a wrong identification, i.e. $P_{k}(\ell_{\tau}=(-1)^{k}a_{k\oplus 1})=\alpha_{k\oplus 1}$, while hitting the boundary $(-1)^{k\oplus 1} a_{k}$
is associated to a successful identification, i.e. $P_{k}(\ell_{\tau}=(-1)^{k\oplus 1}a_{k})=1-\alpha_{k\oplus 1}$. Therefore the expectation value of $\ell$ at the stopping time is approximately (up to the possible overshooting) given by:
\begin{align}
\expect_{k}[\ell(\mathcal{Y}_{\tau})] &= (-)^{k\oplus 1}(-a_{k\oplus 1}\alpha_{k\oplus 1}+a_{k}(1-\alpha_{k\oplus 1})).
\end{align}
Combining the above expressions with $\expect_{k}[\ell_{1}]=(-1)^{k\oplus 1} \mu$ (see \eqref{eq:exl1}) we find
\begin{align}
\label{eq:mtau}
\expect_{k}[\tau_{\mathrm{s}}] =\frac{\expect_{k}[\ell(\mathcal{Y}_{\tau})]}{\expect_{k}[\ell(y)]}=
 \frac{a_{k}(1-\alpha_{k\oplus 1})-a_{k\oplus 1}\alpha_{k\oplus 1}}{\mu}
\end{align}
that, in the regime where the errors (either strong or weak error conditions) are asymptotically small, i.e. $a_{k}  1$, reduces to:
\begin{align}\label{eq:avgstopiid}
\expect_{k}[\tau_{\mathrm{s}}] \sim
 \frac{a_{k}}{\mu}\sim -\frac{\log{\epsilon_{k}}}{\mu}
\end{align}
for both weak and strong error conditions, i.e. for $\mathcal{S}\in \mathbf{C}(\epsilon_{0},\epsilon_{1})$and $\mathcal{S}\in \mathbf{S}(\epsilon_{0},\epsilon_{1})$. Here we used that the SPRT thresholds $a_{k}$ required to meet the weak (strong) conditions, given by equation \eqref{eq:statistial_errors} (equation \eqref{eq:link}) scale as $a_{k}\sim -\log\alpha_{k}$  ($a_{k}\sim -\log\epsilon_{k}$).

Now we are in a position to compare the performances of sequential and deterministic strategies. The time required for a deterministic strategy to guarantee a  small error probability is
given by the asymptotic error rates studied above. For instance, if we need to impose a small error for both hypotheses we have
\be
 T_{\det}\sim \frac{ \log\epsilon}{\min\{R_{0},R_{1}\}}=\frac{ \log\epsilon}{R_{\mathrm{sym}}}=
4 \frac{ \log\epsilon}{\mu}\sim 4\expect_{k}[\tau_{\mathrm{s}}].
\ee
That is the use of sequential strategies allows to save a significant amount of resources:
a deterministic strategy would require 4 times longer sample times to attain the same error bounds than the expected time for the SPRT.  The advantage is also clear in the asymmetric setting: while deterministic strategies are inevitably bound by the trade-off between type I and type II errors (green area in Fig. \ref{fig:ROC}) sequential strategies allow to minimize both error rates simultaneously, up to the absolute non-trivial optimal value given by Stein's lemma (orange area in Fig. \ref{fig:ROC}).

Having introduced the hypothesis testing framework and analyzed the IID Gaussian example in detail, we now shift our focus to the main subject of this work:
 sequential hypothesis testing in continuously-monitoring quantum systems.
  We start by defining the physical systems that we have in mind and derive
  the statistics that will govern the observed signals.

\section{Continuously-monitored Quantum Systems}\label{sec:QMON}
Continuously-monitored quantum systems are systems from which a certain amount of information is extracted at each instant of time. The act of extracting information from the system, or in other words the act of measuring, perturbs the system by an amount that increases with the information extracted from it. At one extreme, under a fully informative measurement,  corresponding to a sharp, or rank-1 POVM, at each given time, the system collapses to the same measured state consistently, preventing any evolution (known as the quantum Zeno effect). At the other extreme,  enforcing no perturbation to the system dynamics will end up with a  completely uninformative measurement.
Hence, in a continuously monitored system, a balance is sought between the amount of information extracted at any one time and the measurement-induced perturbation that can be tolerated on the dynamics. This balance between perturbation and informativeness may be controlled by tuning the coupling of the system of interest with an external environment, on which a sharp measurement is performed.
A Markovian approximation concerning the system dynamics is often assumed, i.e. the environment is assumed to either reset on a very fast time scale or it is assumed to be large enough to guarantee that the information leaked from the system does not kick back. If both the coupling with the bath and measurements are weak, and the observational model is assumed to be Gaussian, then the dynamics of a continuously monitored system is well described by the Belavkin-Zakai equation~\cite{belavkin1989nondemolition, wiseman1993quantum, wiseman2009quantum, jacobs2014quantum}, which here we review:

\begin{align}
\label{eq:BZ}
d \rho_{t} &=
\mathcal{L}_{\theta}(\rho_{t})dt \nonumber\\ 
  &+\left(\mathcal{H}_{\sqrt{\eta} \hat{\mathbf{c}}} (\rho_{t})dt-\Tr(\mathcal{H}_{\sqrt{\eta} \hat{\mathbf{c}}} (\rho_{t})) \rho_{t} \right) \cdot  d\mathbf{w}_{t}
\end{align}
where
\begin{align}
 \mathcal{L}_{\theta}(\rho_{t}) = -i [\hat{H}, \rho_{t}] + \mathcal{D}_{\hat{\mathbf{c}}}(\rho_{t})
 \end{align}
 is the generator of the unconditional dynamics ---in which no measurement is performed, or  equivalently, no measurement record is registered---, $\hat{H}$ is the free Hamiltonian and $\mathcal{D}_{\hat{\mathbf{c}}}(\rho_{t})=\hat{\mathbf{c}}\rho_{t}\hat{\mathbf{c}}^{\dagger}- \frac{1}{2}\{\hat{\mathbf{c}}^{\dagger}\hat{\mathbf{c}},\rho_{t}\}$ is a diffusive term due to the presence of an external bath (where $
\hat{\mathbf{c}}= (\hat{c}_{1}, \dots, \hat{c}_{k})^{T}$ are generic operators defined by the structure of the Markovian environment). The second line of \eqref{eq:BZ} represents the measurement back-action, where
\be\label{eq:inn}
d\mathbf{w}_{t}=  d\mathbf{y}_{t}-\Tr[\mathcal{H}_{\sqrt{\eta} \hat{\mathbf{c}}}(\rho_{t})]dt \ee is the so-called \textit{innovation term}, $d\mathbf{y}_{t}$ is the measurement outcome at the given time $t$, and $\mathcal{H}_{\sqrt{\eta} \hat{\mathbf{c}}}(\sigma_{t})=\sqrt{\eta} \hat{\mathbf{c}}\sigma_{t}+\sigma_{t} \sqrt{\eta} \hat{\mathbf{c}}^{\dagger}$ with $\eta$ a positive semi-definite matrix representing the detection efficiency, which also controls which modes are effectively monitored.

Plugging \eqref{eq:inn} into \eqref{eq:BZ} allows us to write a stochastic equation of motion for the current state of the system conditional
to a particular measurement record $\mathcal{Y}_{t}= (d\mathbf{y_{0}},\dots d\mathbf{y}_{t})$. The non-linear term appearing in the resulting equation describes the re-normalization process of the state after each measurement, i.e. it incorporates the Born rule in the system state dynamics which can be understood as the quantum counterpart of a Bayesian update on the statistical operator.

The probability of having the measurement outcome $d\mathbf{y}_{t}$ at time $t$ is described by the Gaussian observational model:

\begin{align}\label{eq:probdy}
&P(d\mathbf{y}_{t}|\mathcal{Y}_{t},\theta) = \frac{1}{\sqrt{2\pi k dt}}\ex{-\frac{\mathbf{d w}_{t}\cdot  \mathbf{d w}_{t} }{2dt}} \nonumber\\
&\hspace{0.2cm}= \ex{-\frac{1}{2}| \Tr[\mathcal{H}_{\sqrt{\eta}\hat{\mathbf{c}}}(\rho_{t}))]|^{2}dt + \Tr[\mathcal{H}_{\sqrt{\eta} \hat{\mathbf{c}}}(\rho_{t})]\cdot d\mathbf{y}_{t}}P_{d\mathbf{w}}(d\mathbf{y}_{t}),
\end{align}
where in the second equality we have used \eqref{eq:inn}, and $P_{d\mathbf{w}}(d\mathbf{y}_{t})$ is the probability distribution of $k$ independent Wiener processes such that $\mathbb{E}[dw_{i}dw_{j}]= \delta_{ij} dt$.

The probability distribution of the measurement record $\mathcal{Y}_{t}$, conditioned on the system state ---whose dynamics is governed by \eqref{eq:BZ}--- is readily obtained as the product of the probabilities in \eqref{eq:probdy} describing the probability measurement outcome $d\mathbf{y}_{t}$ at a given time and reads
\begin{align}
P(\mathcal{Y}_{t}|\theta)  = \ex{\lambda(\mathcal{Y}_{t}|\theta) } P_{W}(\mathcal{Y}_{t})
\end{align}
where Wiener process $P_{{W}}(\mathcal{Y}_{t}) \equiv  \prod_{\tau=0}^{t} P_{d\mathbf{w}}(dy_{t})$ can be understood as the noise process driving the system, and is what mathematically defines the probability measure in the space of the continuous measurement signals.

Here, $\lambda(\mathbf{y}_{t}|\theta)$ is the log-likelihood associated with the sequence of measurement $\mathcal{Y}_{t}$ and is described by the following master equation~\footnote{The attentive reader may notice the resemblance with the Kallianpur-Striebel log-likelihood~\cite{kallianpur2013stochastic}, describing a non-linear filtering problem. However the term $\lambda(\mathcal{Y}_{t}|\theta)$ must be understood as the log-likelihood of the unnormalized conditional probability density function solving the Duncan-Mortensen-Zakai equation.}
 \begin{align}\label{eq:mela}
d{\lambda}(\mathcal{Y}_{t}|\theta) &=  \Tr(\mathcal{H}_{\sqrt{\eta}\hat{\bm{c}}}(\rho_{\theta}(t,\mathcal{Y}_{t}))\! \cdot\! d\mathbf{y}_{t}\nonumber\\
&\hspace{1cm}-\frac{1}{2} |\Tr(\mathcal{H}_{\sqrt{\eta} \hat{\bm{c}}}(\rho_{\theta}(\mathcal{Y}_{t})|^{2}  dt
 \end{align}
 where, for clarity, we have made the dependence of the state on the parameters $\theta$ and the measurement record $\mathcal{Y}_{t}$ explicit.

The above equation allows us to easily keep track of the likelihood of a particular trajectory $\mathcal{Y}_{t}$, as it can be computed recursively, only knowing its current value, the conditional state of the system, and the measurement outcome at each given time.

Before moving further few remarks are in order.

First notice that \eqref{eq:mela} can also be understood as the equation governing the evolution of trace of the unnormalized state described by the linear Belavkin-Zakai equation, i.e.
\begin{align}\label{eq:sigma}
d \tilde{\rho}_{t} =  \mathcal{L}_{\theta}(\tilde{\rho}_{t})dt + \mathcal{D}_{\sqrt{\eta}\hat{\mathbf{c}}}(\tilde{\rho}_{t})dt +\mathcal{H}_{\sqrt{\eta}\hat{\mathbf{c}}}(\tilde{\rho}_{t})d\mathbf{y}_{t},
\end{align}
as  {firstly shown in~\cite{Tsang2012Continuous} and subsequently rediscovered in~\cite{gammelmark2013bayesian}}~\footnote{There is a discrepancy of a factor 1/2 in the second term of the r.h.s. of \eqref{eq:mela} with respect to  (26) in~\cite{gammelmark2013bayesian}. However, it is not difficult to show that, starting from (26) in~\cite{kiilerich_hypothesis_2018} and with the help of the It\^{o} calculus the correct eq.~\eqref{eq:mela} is obtained. }.
This fact should not be surprising, since the linear Belavkin-Zakai equation can also be understood as the quantum counterpart of the classical Duncan-Mortensen-Zakai equation~\cite{duncan1967probability}, i.e. the equation describing the dynamics of the unnormalized conditional probability density function for a classical non-linear filtering problem~\cite{duncan1967probability, mortensen1966optimal}.

The above equations are usually hard to handle, especially in the case of continuous, infinite-dimensional systems. Even conducting numerical simulations
 often demands extensive allocation of computational resources.
However, there is a full class of experimentally relevant systems whose behavior can be approximated by the evolution of a Gaussian state. Here, the system is described by the first two statistical moments of the quadratures, and its evolution is fully characterized by a system of linear stochastic differential equations. Such Gaussian models are employed to describe several real-world scenarios with great success~\cite{Aspelmeyer2014Cavity,delic2020cooling1, Rossi2020Experimental}.
They are not only a useful tool to describe real experimental scenarios but they can also be employed to have a more transparent connection between continuously monitored systems dynamics and classical filtering theory~\cite{Doherty1999, trees_detection_2001}.

\subsection{ Gaussian systems}\label{ssec:QMONgauss}

A quantum Gaussian system of $n$-modes is described by the quadratures $\mathbf{q}$ and $\mathbf{p}$ with $[q_{i},p_{j}]= i \delta_{ij}$, whose unitary part of the dynamics is described by a quadratic Hamiltonian of the form $\hat{H}= \frac{1}{2}\hat{\mathbf{x}}^{T}H\hat{\mathbf{x}}+\mathbf{b}^{T} \Omega \hat{\mathbf{x}}$, where $\mathbf{x} = (\hat{q}_{1}, \hat{p}_{1}, \hat{q}_{2}, \hat{p}_{2}, . . . ,\hat{q}_{n}, \hat{p}_{n})^{T} $, $H$ is a $2n \times 2n$-matrix, $\mathbf{b}$ is a $2n$-dimensional vector accounting for a time-dependent linear driving, and $\Omega$ is the $n$-mode symplectic matrix~\cite{serafini2017,weedbrock2012Gaussian}.
The effect of the environment is described by Lindblad generators that are linear in the system's quadratures, as well as the noisy measurement which is described by a linear function of the quadratures.
So that the dynamics will preserve the gaussianity of the state~\cite{genoni2016conditional}. This means that if also the initial state is Gaussian the first moment $\mathbf{r}= \braket{\hat{\mathbf{r}}_{t}} $ and the Covariance Matrix (CM) $\mathbf{\sigma}=  \braket{\{\hat{\mathbf{r}}_{t},\hat{\mathbf{r}}_{t} \}}/2- \braket{\hat{\mathbf{r}}_{t}}\braket{\hat{\mathbf{r}}_{t}}$ are enough to characterize the system at any time. The evolution for the first two momenta is thus obtained through \eqref{eq:BZ} and reads
\begin{align}\label{eq:LSE}
d\mathbf{r}_{t} &= A_{\theta} \mathbf{r}_{t} dt +  \mathbf{b}_{\theta,t}dt +\chi(\sigma_{t})(d\mathbf{y}_{t}- C \mathbf{r}_{t} dt)\nonumber\\
\dot{\sigma_{t}} & = A_{\theta} \sigma_{t} +\sigma_{t} A_{\theta}^{T} + D_\theta - \chi(\sigma_{t}) \chi(\sigma_{t})^T,
\end{align}
and the model describing the measured signal simplifies to
\begin{align}
\label{eq:dy}
    d\mathbf{y}_t = C \mathbf{r}_{t} dt + d\mathbf{w}_t.
\end{align}
where $A_{\theta}$ is the drift matrix and takes into account the unitary interaction between the system and environment, as well as the internal dynamics of the system, $\mathbf{b}_{\theta,t}$ describes the effects of a (possibly time-dependent) force on the system and $D_\theta$ the diffusive part of the dynamics due to the interaction with an environment, and $\chi(\sigma):= \sigma C^T -\Gamma$ accounts for the measurement back-action.
The sub-index $\theta$ denotes the dependence on certain parameters that characterize the different hypotheses.

 Note that on one hand, the dynamics of the first moment is perturbed by the measurement back-action, of an amount proportional to the innovation term, $d\mathbf{w}_{t}$, and hence it explicitly depends on the measured signal. While, on the other hand, the dynamics of variance ---while influenced by the measurement process--- does not depend on the particular measured signal but is reduced by an amount proportional to the averaged fluctuation of the first moment induced by the measured signal.

 It is furthermore worth noticing that \eqref{eq:LSE} are mathematically equivalent to the Kalman-Bucy equations~\cite{Doherty1999} solving the classical filtering problem of estimating the internal state in a linear dynamics system from a series of noisy measurements~\cite{bucy1961new}.  The quantum formalism effectively includes both the Bayesian (state of knowledge changes) and the quantum measurement back-action. In turn, eqs.~\eqref{eq:LSE} can be understood as those governing the evolution of the first two moments of the normalized probability distribution obtained through the Bayesian update of a classical linear system under the information acquired through the sequence of linear noisy measurements~\cite{Doherty1999}.

Having discussed the quantum continuously monitoring systems, we now apply the statistical inference tools discussed in Sec.~\ref{sec:hypothesis_testing} to this kind of systems.

\section{Sequential Hypothesis testing in continuously-monitored systems}\label{sec:QMONtest}
Now that the basic notions and formalism have been set, let us discuss  {sequential} hypothesis testing in a quantum system that is being continuously monitored.
Contrary to previous studies of  {in quantum sequential analysis} \cite{QuantumVargas2021,fanizza_ultimate_2023}, where copies of a quantum state are provided on demand, here one has a single system that evolves due to its own internal dynamics and the effects of measurements.
The measurement record $\mathcal{Y}_{t}$ cannot be described via a sequence of IID but is instead generated by a specific hidden Quantum Markov Model.
 In any case, what it is clear from the previous section is  { that the central quantity} to be studied is the LLR.
\begin{align}
\ell(\mathcal{Y}_{t}) = \log \frac{p_{1}(\mathcal{Y}_{t})}{p_{0}(\mathcal{Y}_{t})}.
\end{align}
Exploiting \eqref{eq:mela} it is  {immediate} to show that $\ell({\mathcal{Y}_{t}})$ is characterized by the following differential equation:
\begin{align}\label{eq:sqrt_log}
&d \ell({\mathcal{Y}_{t}}) =  \Tr[\mathcal{H}_{\sqrt{\eta}\hat{\mathbf{c}}}(\rho_{\theta_{1}}(\mathcal{Y}_{t})-\rho_{\theta_{0}}(\mathcal{Y}_{t})]\cdot d\mathbf{y}_{t}\nonumber\\
&\quad\, -\frac{1}{2} (|\Tr[\mathcal{H}_{\sqrt{\eta}\hat{ \mathbf{c}}}\rho_{\theta_{1}}(\mathcal{Y}_{t}]|^{2} -|\Tr[\mathcal{H}_{\sqrt{\eta}\hat{\mathbf{c}}}\rho_{\theta_{0}}(\mathcal{Y}_{t}]|^{2})dt,
\end{align}
where $\rho_{\theta_{0}}(\rho_{\theta_{1}})$ is determined by solving \eqref{eq:BZ} under the null(alternative) hypothesis.
 { This expression was derived by Tsang in~\cite{Tsang2012Continuous}  where (deterministic) hypothesis testing in continuously monitored quantum systems was discussed for the first time.}

It is worth noticing that~\eqref{eq:sqrt_log} can be further simplified assuming the sequence of outcomes $\mathcal{Y}_{t}$ is generated through the null/alternative hypothesis, denoted by $k$ (i.e. one of the models correctly describes the system's dynamics). Indeed, under this assumption we can express $d\textbf{y}_{t}$ as
\begin{align}
d\mathbf{y}_{t} \equiv d\mathbf{y}_{t|\theta_{k}} \equiv \Tr[\mathcal{H}_{\sqrt{\eta}\hat{ \mathbf{c}}}\rho_{\theta_{k}}(\mathcal{Y}_{t|\theta_{k}})]dt + d\mathbf{w}_{t},
\end{align}
and rewrite the LLR $\ell(\mathcal{Y}_{t|\theta_{k}})$ as a function of the Wiener noise $d\mathbf{w}_{t}$,
\begin{align}\label{eq:ellk}
&d \ell(\mathcal{Y}_{t|\theta_{k}}) =
\Tr[\mathcal{H}_{\sqrt{\eta} \hat{ \mathbf{c}}}(\rho_{\theta_{1}}(\mathcal{Y}_{t|\theta_{k}})-\rho_{\theta_{0}}(\mathcal{Y}_{t|\theta_{k}}))]\cdot d\mathbf{w}_{t}\nonumber\\
&\hspace{0.7cm}(-)^{k\oplus 1} \frac{1}{2}|\Tr[\mathcal{H}_{\sqrt{\eta} \hat{\mathbf{c}}}(\rho_{\theta_{1}}(\mathcal{Y}_{t|\theta_{k}})-\rho_{\theta_{0}}(\mathcal{Y}_{t|\theta_{k}}))]|^2 dt.
\end{align}
In the Gaussian case, the equation for the LLR is:
\begin{align}\label{eq:linealLogy}
    d \ell_t &=   (C \Delta\mathbf{r}_{t}) \cdot d\mathbf{y}_{t}-\frac{1}{2}
    (||C \mathbf{r}_{1}(t)||^{2}-||C \mathbf{r}_{0}(t)||^{2})dt,
\end{align}
where $\mathbf{r}_{k}(t)$ are the first moments of the Gaussian state generated through
the conditional dynamics of \eqref{eq:LSE} fed with  $\mathcal{Y}_{t|\theta_{k}}$,
and $\Delta\mathbf{r}_{t} = \mathbf{r}_{1}(t)- \mathbf{r}_{0}(t)$.

Similarly, in terms of the innovations or Wiener noises:
\begin{equation}\label{eq:linealLog}
    d \ell_t=  \frac{(-1)^{k\oplus 1}}{2}||C \Delta\mathbf{r}_{t}  ||^2 dt + (C \Delta\mathbf{r}_{t}) \cdot d\mathbf{w}_t.
\end{equation}

The above expression shows that the likelihood ratio has a positive or negative drift depending on whether the measured signal $\mathcal{Y}_{t}$ is generated by the alternative or null hypothesis, showing the tendency to increase the chances to discriminate between the two hypotheses over time correctly.
Notice that the same conclusion can be obtained, in full generality by means of the following identity:
$\frac{P_{1}(\ell_{t})}{P_{0}(\ell_{t})}=\ex{\ell_{t}}$.
Indeed $\expect_{k}[\ex{(-)^{k}\ell_{t}}] =1$ and with the help of Jensen inequality, i.e. $\ex{\expect_{k}[{(-)^{k}\ell_{t}}]} \le \expect_{k}[\ex{(-)^{k}\ell_{t}}]$ one readily obtains
\begin{align}
(-)^{k}\expect_{k}[\ell(\mathcal{Y}_{t})]\le0.
\end{align}
We note that equation \eqref{eq:sqrt_log} allows for immediate implementation of an SPRT in a real experiment, while \eqref{eq:ellk} is more suited to theoretically study the performance of the SPRT in the context of continuously monitored systems as we will see next.

We observe that integrating \eqref{eq:linealLog} would result in a Gaussian distributed LLR (compare also with \eqref{eq:ellGauss}) if $\Delta\mathbf{r}$ were a constant. Such a scenario would greatly simplify our analysis. Unfortunately, this quantity is far from being constant as it is a stochastic variable itself, strongly correlated with the noise process.
The next section is devoted to presenting the main formal results of this paper, where under some mild assumptions on the statistical properties of the LLR we derive the statistics of the most relevant quantity in sequential methodologies, namely the stopping time $\tau$.

\begin{figure}[htb!]
\centering
\includegraphics[width=0.45\textwidth]{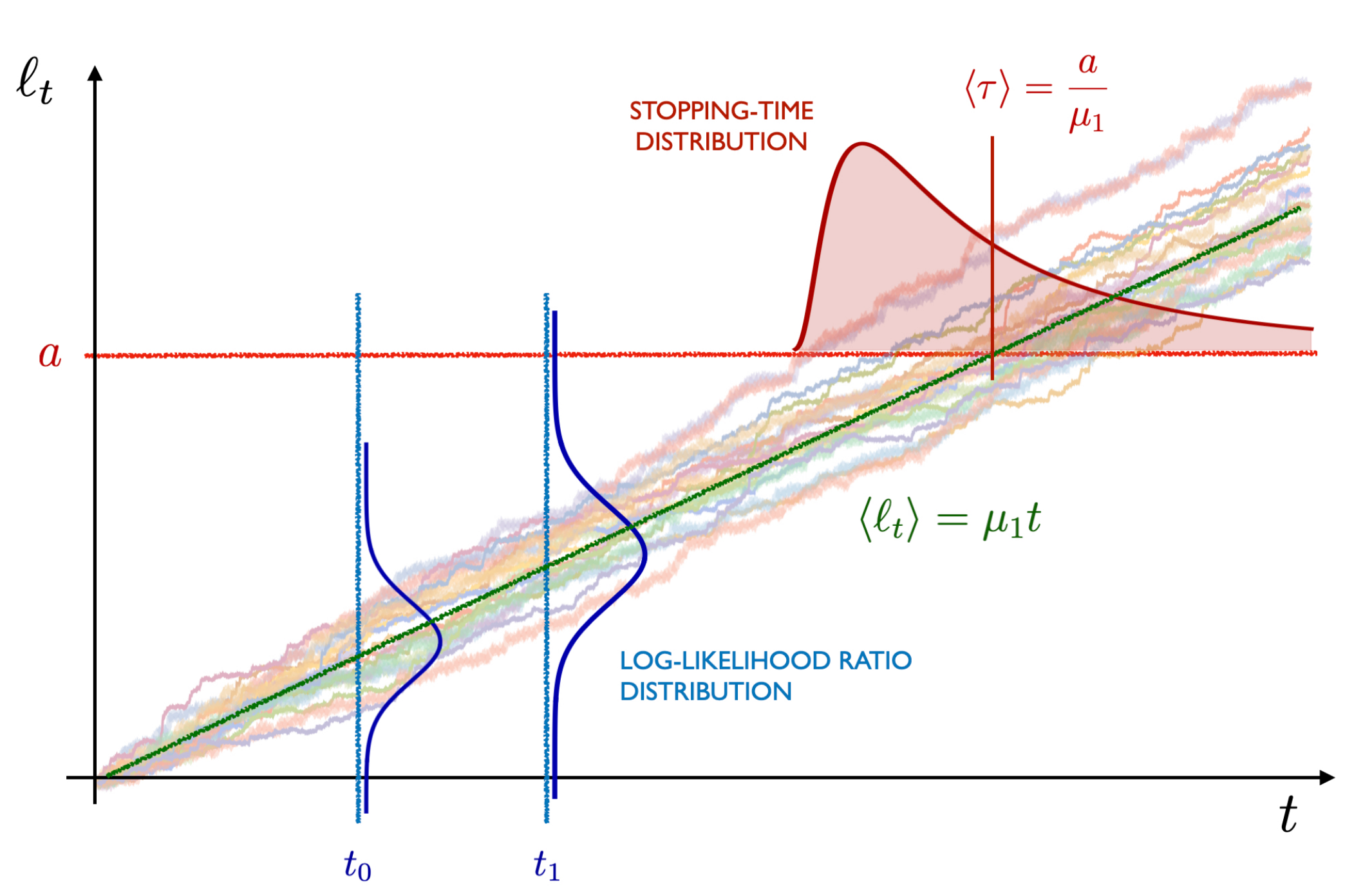}
\caption{\textit{We illustrate how the stopping time distribution arises from
    the stochastic continuous trajectories of $\ell_{t}$; here we show some realizations of the $\ell$ process, along with some time-slices at $t_0$ and $t_1$ shown in blue together with the corresponding distributions $p_{1}(\ell_t)$, which close to their peak value are well approximated by a Gaussian. The horizontal line in red corresponds to the fixed threshold $\ell_{\tau}=a$, leading to an arrival time distribution $p_{1}(\tau)$. Such distribution appears as a consequence of a difference in the arrival times for trajectories of $\ell_{t}$.}}
\label{fig:LLRfig}
\end{figure}

Before that, let us conclude this section with Figure \ref{fig:LLRfig} illustrating the SPRT strategy in continuously monitored systems where we anticipate some of the general features that we will formally prove in the next section.

\subsection{General results on Discrimination in Quantum Continuously Monitored Systems}\label{ssec:theorems}
We now present an analytical study of the performance of the SPRT in hypothesis testing for continuously monitored quantum systems.
We give a general theorem providing a tool to upper bound the SPRT stopping time and show the optimality of the test under some assumptions on the underlying stochastic process, and derive some relevant statistical properties of the SPRT stopping time. We further provide
the optimal rate of error for the asymmetric deterministic setting, i.e. the Stein lemma for continuously monitored systems.  All the details and the proofs of the following theorems can be found in the Supplemental Material.

Along with the asymptotic analysis notation introduced in previous sections (see  \cref{footAs}) let us introduce the notation for convergence in probability:  Let $X_{t}$, $Y_{t}$ be random sequences taking values in any normed space, we use the compact notation $ X_{t}= \smallO_{p}(Y_t)$ to denote
$$\lim_{t \to \infty} P\left(||{X_t}||> \epsilon ||Y_{t}|| \right)= 0 \hspace{1.5cm} \forall \epsilon > 0.$$
\begin{theorem}\label{th:optepsilon}

\textit{(SPRT optimality)}. Let $\mathcal{S}=(d,\tau)$ denote a generic hypothesis test where $\tau$ is a stopping time, $d=d(\mathcal{Y}_{\tau})$  is a terminal decision function with values in the set $\{0,1\}$, and $\mathcal{Y}_{\tau}$ the full samples acquired in the time interval $[0,\tau]$.
Let $\mathbf{C}(\alpha_{0},\alpha_{1})=\{\mathcal{S} : P_{0}(d=1)\le \alpha_{0},P_{1}(d=0)\le \alpha_{1}\}$ and $\ell_{t}\equiv \log \frac{P_{1}(\mathcal{Y}_{t})}{P_{0}(\mathcal{Y}_{t})}$ the LLR. The SPRT is defined by the couple $\mathcal{S}_{\mathrm{s}} = (d_{\mathrm{s}},\tau_{\mathrm{s}})$, where
\begin{align}
 \tau_{\mathrm{s}} = \inf\{t \ge0 ; \ell_{t}\notin (-a_{0},a_{1})\}
\end{align}
and
\begin{align}
    d_{\mathrm{s}}=\begin{cases}
    1 &\textit{if}\quad \ell_{\tau_{\mathrm{s}}}\ge a_{1}\\
    0 & \textit{if}\quad \ell_{\tau_{\mathrm{s}}}\le -a_{0}\\
\end{cases}
\end{align}
with $a_{k}>0$ for $k=0,1$.
Let $T$ be a generic time and $\alpha^{*}=\max\{\alpha_{0},\alpha_{1}\}$ then
\begin{align}
    \lim_{\alpha^{*}\to 0 }P_{k}(\tau\ge T) \ge \lim_{\alpha^{*}\to 0 } P_{k}(\tau_{\mathrm{s}}^{(\delta)}\ge T)
\end{align}
and
\begin{align}
    P_{k}(\tau\ge \tau_{\mathrm{s}}^{(\epsilon)}) = 1-\mathcal{O}(\alpha_{k\oplus 1}^{\varepsilon}),
\end{align}
where $\tau$ and $\tau_{\mathrm{s}}^{(\delta)}$ respectively denote the stopping time of a generic hypothesis test in the class $\mathcal{S} \in  \mathbf{C}(\alpha_{0},\alpha_{1})$ and the stopping time of the SPRT in the class $\mathcal{S}_{\delta} \in  \mathbf{C}(\alpha_{0}^{1+\delta},\alpha_{1}^{1+\delta})$ with $\epsilon \in (0,1)$, and $\varepsilon= \frac{\delta}{1+\delta}$.
\end{theorem}
This first theorem is fairly general and allows us to asymptotically lower-bound the stopping time of a generic test in the class $\mathbf{C}(\alpha_{0},\alpha_{1})$, by that associated to an SPRT. A less abstract result and an asymptotic optimality condition can be obtained if some reasonable assumptions over the SPRT stopping time are made.

\begin{corollary}\label{cor:weakopt}
If, under the hypothesis $h_{k}$,
\begin{align}\label{eq:tauconvass}
\tau_{\mathrm{s}} = \frac{\log(\alpha_{k})}{\mu_{k}}+\smallO_{p_{k}}(\log(\alpha_{k}))
\end{align}
with $\tau_{\mathrm{s}}$ the SPRT stopping time in the class $\mathbf{C}(\alpha_{0},\alpha_{1})$, then the SPRT is asymptotically optimal in a weak sense, i.e.
\begin{align}\label{eq:weakoptim}
\lim_{\alpha^{*}\to 0}  P_{k}(\tau\ge (1-\epsilon)\tau_{s})=1\hspace{1cm} \forall\epsilon>0
\end{align}
where $\alpha^{*}= \max(\alpha_{0},\alpha_{1})$.
\end{corollary}
\begin{corollary}
If furthermore
\begin{align}
  \expect_{k}[\tau_{s}^{n}] = \left(\frac{\log(\alpha_{k})}{\mu_{k}}\right)^{n}+\smallO(\log(\alpha_{k})^n)
\end{align}
the SPRT is not only asymptotically optimal in the weak sense but also in momenta, i.e.
\begin{align}\label{eq:asymptsuppl2}
\expect_{k}[\tau^{n}]\ge \expect_{k}[\tau_{\mathrm{s}}^{n}](1+\smallO(1)).
\end{align}
\end{corollary}
On the other hand, the following theorem sets a bound on the optimal error rate that the (asymmetric) deterministic setting can achieve:
\begin{theorem}\label{th:4}(Continuously-monitoring Stein lemma). Let $P_{k}$ be the probability under the hypothesis $h_{k}$ with $k=0,1$, $d= d(\mathcal{Y}_{T})$ a decision function in the set $\{0,1\}$, and $T$ a fixed time at which the decision is taken. Let $\alpha_{0}$ and $\alpha_{1}$ respectively be the type I and II errors,
\begin{align}
\alpha_{k}(T)^{*} := \min\{\alpha_{k}(T) : \alpha_{k\oplus 1}(T) \le \epsilon \}
\end{align}
with $\epsilon \in (0,1)$, i.e. the minimum error achievable in the asymmetric scenario and
$$R_{k}^{*}:= \lim_{T\to \infty} -\frac{\log \alpha^{*}(T)}{T}$$
the corresponding error rate.
If
\begin{align}
\ell(\mathcal{Y}_{t}) = (-)^{k\oplus 1}\mu_{k} t+\smallO_{p_{k}}(t)
\end{align}
with $\mu_{k}> 0$, then the minimum error rate $R_{k}$ that can be attained by a deterministic test is given by the following equation
\begin{align}
R_{k}^{*} = \mu_{k}.
\end{align}
\end{theorem}
The proof of the theorem can be found in the supplemental (Theorem~\ref{sec:optima_asym}).
As shown in the Supplemental Material, this is indeed the faster error rate we can achieve using a deterministic strategy, in the sense that any fastest rate will lead with certainty to false positive (negative), i.e. $\alpha_{0(1)}\to 1$. Also notice that if $\ell_{t}$ converges in mean, then $\mu_{k}$ assumes the role of the regularized Kullback-Leibler information divergence, i.e.
\begin{align*}
\mu_{k}=I(P_{k}\| P_{k\oplus 1}) \equiv\lim_{t \to \infty} \frac{1}{t} \expect_{k}\left[\log \frac{P_{k}(\mathcal{Y}_{t})}{P_{k\oplus 1}(\mathcal{Y}_{t})}\right].
\end{align*}

\begin{theorem}\label{th:1}
Let $\ell(\mathcal{Y}_{t})$ be the LLR described by \eqref{eq:sqrt_log} and $\tau_{\mathrm{s}}$ be a stopping time associated to the SPRT with thresholds $a_{k}$. If
\begin{align}
\ell(\mathcal{Y}_{t}) = (-)^{k\oplus 1}\mu_{k} t +\smallO_{p_{k}}(t)
\end{align}

with $\mu_{k}> 0$ then, on the one hand,
 limit of small error bounds  (i.e. $a_{k}  1$)  the stopping time

 converges in probability to a constant:
\begin{align}
\tau_{\mathrm{s}} = \frac{a_{k}}{\mu_{k}}+\smallO_{p_{k}}(a_{k})
\end{align}
where recall that $a_{k}\sim -\log(\epsilon_{k})$, as $\max(\epsilon_{0},\epsilon_{1})\to 0$ (or similarly for the weak error conditions replacing $\epsilon$ by $\alpha$).

On the other hand, under \textit{weak error} conditions the SPRT is asymptotically optimal in a weak sense, i.e.
\begin{align}
\lim_{\alpha^{*}\to 0}P(\tau \ge (1-\varepsilon)\tau_{\mathrm{s}})=1 \; \; \forall \;\varepsilon\in (0,1)
\end{align}
with $\alpha^{*}=\max(\alpha_{0},\alpha_{1})$, and $\tau$ the stopping time of a generic test in the class of $\mathbf{C}(\alpha_{0},\alpha_{1})$.

\begin{corollary}
If we further assume
\begin{align}
\expect_{k}[\ell_{t}] = (-)^{k\oplus 1}\mu_{k}t +\smallO(t)
\end{align}
then
 the (asymptotic) mean stopping time is given by
\begin{align}
\expect_{k}[\tau_{s}] = \frac{a_{k}}{\mu_{k}}+\smallO(a_{k})
\end{align}
while for \textit{weak error} conditions the SPRT is asymptotically optimal in mean, i.e.
\begin{align}
\expect_{k}[\tau] \ge \expect_{k}[\tau_{s}]+\smallO(\log(\alpha_{k}))
\end{align}
for $\tau$ and $\tau_{s}$ in the class $\mathbf{C}(\alpha_{0},\alpha_{1})$.
\end{corollary}

\end{theorem}
\begin{theorem}\label{th:43}
Let $\ell(\mathcal{Y}_{t})$ be the LLR described by \eqref{eq:sqrt_log} and $\tau_{\mathrm{s}}$ be the stopping time associated to the SPRT.
If
\begin{align}\label{eq:ellgauss}
\ell(\mathcal{Y}_{t}) = (-)^{k\oplus 1} \mu_{k} t +\sigma_{k}\zeta_{t}+ \smallO_{p_{k}}(\sqrt{t})
\end{align}
with $\mu_{k}\in \mathbb{R}^{+}$ and $\zeta_{t}$ a standard Wiener process, i.e. $\mathbb{E}[\zeta_{t}]=0$ and  $\mathbb{E}[\zeta_{t}\zeta_{\tau}]=\min(t,\tau)$,
then the probability distribution of the normalized stopping time $\tilde\tau:=\frac{\tau_{\mathrm{s}}}{a_{k}}$ can be asymptotically approximated by the Inverse Gaussian distribution:
\begin{align}\label{eq:STdist}
p_{k}(\tilde\tau) \sim \frac{a_{k}}{\tilde\tau^{3/2}\sqrt{2\pi \sigma_{k}^{2}a_{k}}}\ex{-\frac{(1-\mu_{k} \tilde\tau)^{2}}{2\sigma_{k}^{2}a_{k}\tilde\tau}}
\end{align}
and the SPRT is \textit{asymptotically optimal in momenta}, i.e.
\begin{align}
\lim_{\alpha^{*}\to 0} \mathbb{E}[\tau^n]\ge \lim_{\alpha^{*}\to 0} \mathbb{E}[\tau^n_{\mathrm{s}}](1-\smallO(1)),
\end{align}
where $\alpha^{*}=\max(\alpha_{0},\alpha_{1})$ and $\tau$ is the stopping time of a generic test in $\mathbf{C}(\alpha_{0},\alpha_{1})$.
\end{theorem}
The proof can be found in the supplemental (Theorem~\ref{sec:optminsupp} and subsequent corollaries).

\subsubsection*{Continuously-monitored gaussian systems}\label{ssec:QMONGAUSStest}

We recall that under the Gaussian assumption, the dynamics is fully characterized by the evolution of the first two cumulants, and reads
\begin{align}\label{eq:LSEa}
d\mathbf{r}_{t} &= A_{\theta} \mathbf{r}_{t} dt +\mathbf{b}_{\theta,t}+\chi(\sigma_{t})(d\mathbf{y}_{t}- C \mathbf{r}_{t} dt)\nonumber\\
\dot{\sigma_{t}} & = A_{\theta} \sigma_{t} +\sigma_{t} A_{\theta}^{T} + D_\theta - \chi(\sigma_{t}) \chi(\sigma_{t})^T.
\end{align}
By picking the parameters $\theta$ corresponding to hypothesis $h_{0}$ and $h_{1}$ respectively, we can write two decoupled sets of equations describing the quantum state of the system conditional on an arbitrary measurement signal $\mathcal{Y}_{t}$ for each of the candidate hypothesis. This suffices to keep track of the LLR and implement the SPRT algorithm.

However, in order to assess its performance and compute the statistical properties of the LLR and the stopping time $\tau_{s}$, it is important to take into consideration the statistical properties of the true measurement signal, which will be governed by \eqref{eq:dy} with $\mathbf{r}_{t}$ corresponding to the true hypothesis. Since this dependence is fed in the stochastic equation for the other (false) hypothesis it results in a coupled system of equations.
To solve this system of equations it is convenient to treat the problem in an extended vector space where the state of the system is defined as $\mathbf{X}_{t}=(\mathbf{r}_{0}(t),\mathbf{r}_{1}(t))^{T}$ and $\Sigma_{t} =\sigma_{0}(t)\oplus\sigma_{1}(t)$ whose evolution is given by

\begin{align}
d\mathbf{X}_{t} &=  (\mathcal{A} -\chi(\Sigma_{t}) \Pi_{k} \mathcal{C}) {\mathbf{X}}_{t} dt + \mathbf{B}_{t}dt + \chi(\Sigma) d\mathbf{w}_{t}\nonumber\\
\dot{\Sigma_{t}} &= \mathcal{A} \Sigma_{t}+\Sigma_{t} \mathcal{A} +\mathcal{D}-\chi(\Sigma_{t})\chi(\Sigma_{t})^{T}
\end{align}
with $\mathbf{B}_{t}= (\mathbf{b}_{0,t},\mathbf{b}_{1,t})^{T}$, $\mathcal{A}=A_{0}\oplus A_{1}$, $\mathcal{C}=C\oplus C$ and $\mathcal{D}= D_{0}\oplus D_{1}$, $\chi(\Sigma_{t}) := \Sigma_{t} \mathcal{C}^T -\tilde{\Gamma}$, $\tilde{\Gamma}= \Gamma_{0}\oplus\Gamma_{1}$  and
\begin{align}
  \Pi_{0}=\begin{pmatrix}
    0 & 0\\
  -\one & \one
   \end{pmatrix},
   \hspace{0,5cm}
   \Pi_{1}=\begin{pmatrix}
  -\one & \one\\
    0   &   0
   \end{pmatrix}
  \end{align}
where $k\in\{0,1\}$ and denotes the hypothesis under which the signal $\mathcal{Y}_{t}$ is generated.
In this notation, the LLR reads:
\begin{align}
d\ell(\mathcal{Y}_{t|\theta_{k}}) =\frac{(-)^{k\oplus 1}}{2} |\mathbb{\Delta}^{T}\mathcal{C}{\mathbf{X}_{t}} |^2 dt + (\mathbb{\Delta}^{T}\mathcal{C}{\mathbf{X}_{t}})\cdot d\mathbf{w}_{t}
\end{align}
with $\mathbb{\Delta^{T}}= (\one,-\one)^{T}$.

Under the assumption that the covariance $\Sigma_{t}$, admits an asymptotic steady state\footnote{Notice that the dynamics of $\Sigma_{t}$ is described by a Riccati equation that we know admits asymptotic steady states in certain regimes.} $\Sigma_{st}= \sigma_{0}\oplus\sigma_{1}$, $\Re[\mathcal{A}-\chi(\Sigma_{st})\Pi_{k}\mathcal{C}] < 0$ and $\mathbf{B}_{t}=\mathbf{B}$, i.e. the affine term of the equation is constant, then the probability distribution of $\mathbf{X}_{t}$ admits an asymptotic solution of the form~\cite{risken1996fokker}:
\begin{align}
  P_{\infty,k}(\mathbf{X})=  \frac{1}{(2\pi)^{n/2} \text{det}[\boldsymbol{\omega}]^{1/2}}\ex{-\frac{1}{2}
  ( \mathbf{X}-\mathbf{d})^{T}\boldsymbol{\omega}^{-1}( \mathbf{X}- \mathbf{d})}
  \end{align}
where $ \mathbf{d}= [\mathcal{A}-\chi(\Sigma_{\infty}\Pi_{k}\mathcal{C})]^{-1}\mathbf{B}$ and $\boldsymbol{\omega}$ is the solution of the Lyapunov equation $(\mathcal{A}-\chi(\Sigma_{t})\Pi_{k}\mathcal{C})\boldsymbol{\omega}+\boldsymbol{\omega}(\mathcal{A}-\chi(\Sigma_{t})\Pi_{k}\mathcal{C})^{T}=2\Sigma_{\infty}$.
From this it is easy to see that:
\begin{align}\label{eq:dlimit}
&\lim_{t\to \infty }\frac{\expect_{k}[d\ell(\mathcal{Y}_{t})]}{dt} = \frac{(-)^{k\oplus 1}}{2}\E [ |\mathbb{\Delta}^{T}\mathcal{C}{\mathbf{X}}|^{2}] \nonumber\\
&= \frac{(-)^{k\oplus 1}}{2}\Tr[\mathcal{C}^{T}\mathbb{\Delta\Delta}^{T}\mathcal{C}\tilde{\boldsymbol{\omega}}]=: \mu_{k}
\end{align}
with $\boldsymbol{\omega}=(\boldsymbol{\omega} +\mathbf{d}\mathbf{d}^{T})$.
 Now, following a similar procedure than in the proof of Wald's identity \cite{wald2004sequential} we have that in the limit of large $a_{k}$'s (i.e. large stopping times)
 \begin{align}
&\expect_{k}\left[\int_{0}^{\tau}d\ell_{t}\right] =
 \expect_k\left[\int_{0}^{\infty} \mathcal{I}_{t<\tau} d\ell_{t}\right]=\nonumber\\
&= \int_{0}^{\infty} \expect_k[\mathcal{I}_{t<\tau}]\expect_k[d\ell_{t}]
 = \mu_{k} \int_{0}^{\infty} \expect_k[\mathcal{I}_{t<\tau}] dt+\nonumber\\
 &\hspace{1em}
 +\int_{0}^{\infty} \expect_k[\mathcal{I}_{t<\tau}](\expect_k[d\ell_{t}]-\mu_{k}dt)\nonumber=\nonumber\\
&=\mu_{k} \expect_k\left[\int_{0}^{\infty} \mathcal{I}_{t<\tau} dt\right]+\mathcal{O}(1)=\nonumber\\&
=\mu_{k} \expect_k\left[\int_{0}^{\tau} dt\right]+\mathcal{O}(1)=\nonumber\\
 &=\mu_{k}\expect_k[\tau]+\mathcal{O}(1),
 \end{align}
 where we have defined the indicator function $\mathcal{I}_{C}=1$ if condition $C$ is fulfilled  and  $\mathcal{I}_{C}=0$ otherwise.
 In the third equality, we have used that $\mathcal{I}_{t<\tau}=
 \mathcal{I}_{\ell_{t}\notin (-a_0,a_1)}$ and $d\ell_{t}$ are independent stochastic variables. Finally, in the fourth equality, we have used that $\expect_k[d\ell_{t}]$ is bounded and
 becomes constant $\mu_{k}$ at a fast enough rate, so as to guarantee that the integral $\int_{0}^{\infty}(\expect_k[d\ell_{t}]-\mu_{k}dt)$ converges to a constant.

 Finally, as in the IID case, we can use the fact that $\ell_{\tau}=\{-a_{0},a_{1}\}$ is a binary random variable (continuity of $\ell_{t}$ guarantees that there's no overshooting) and proceeding along the same lines as in   \eqref{eq:mtau} we obtain an asymptotic expression for the average stopping time:
\begin{align}\label{eq:wald_gaussian}
\expect_{k}[\tau_{\mathrm{s}}] = \frac{a_{k} }{\Tr[{\mathcal{C}^{T}\mathbb{\Delta\Delta}^{T}\mathcal{C}\tilde{\boldsymbol{\omega}}}]}(1+{o}(1)),
\end{align}
where recall that $a_{k}\sim\log\alpha_{k}^{-1}$ ($a_{k}\sim\log\epsilon_{k}^{-1}$) for the strong (weak) error conditions.

If, furthermore the variance of the LLR $\mathbb{E}[\ell_{t}^{2}]-\mathbb{E}[\ell_{t}]^2= \mu_{k}t+o(t)$, with $\mu_{k}<\infty$, then the SPRT is also asymptotically optimal in mean, i.e.
\begin{align}
\mathbb{E}[\tau] \ge \mathbb{E}[\tau_{s}]+\mathcal{O}(1)
\end{align}
where $\tau$ is the stopping time of a generic test in the class $\mathbf{C}(\alpha_{0},\alpha_{1})$, in accordance to Theorem~\eqref{th:43}.

To conclude this work in the next section we will numerically investigate the behavior of the LLR and tests' performances in a specific model, and explain the results in light of the theoretical results obtained so far.

\subsection{Optomechanical Sensors}~\label{sec:linear_opto}
In this section, we study the performance of sequential hypothesis testing on an optomechanical system under homodyne measurement that operates in the linear regime.
We also assume that the system operates in the unresolved sideband regime, which enforces a separation of the time scales of the optical and mechanical modes, allowing the cavity mode to be eliminated adiabatically~\cite{Doherty1999}.
In this regime, the fluctuations evolve according to a quadratic Hamiltonian, ensuing a Gaussian evolution described by \eqref{eq:LSE}, where the number of modes is reduced to one, i.e. the mechanical mode.

\begin{figure}[t!]
\centering
\includegraphics[width=0.5\textwidth]{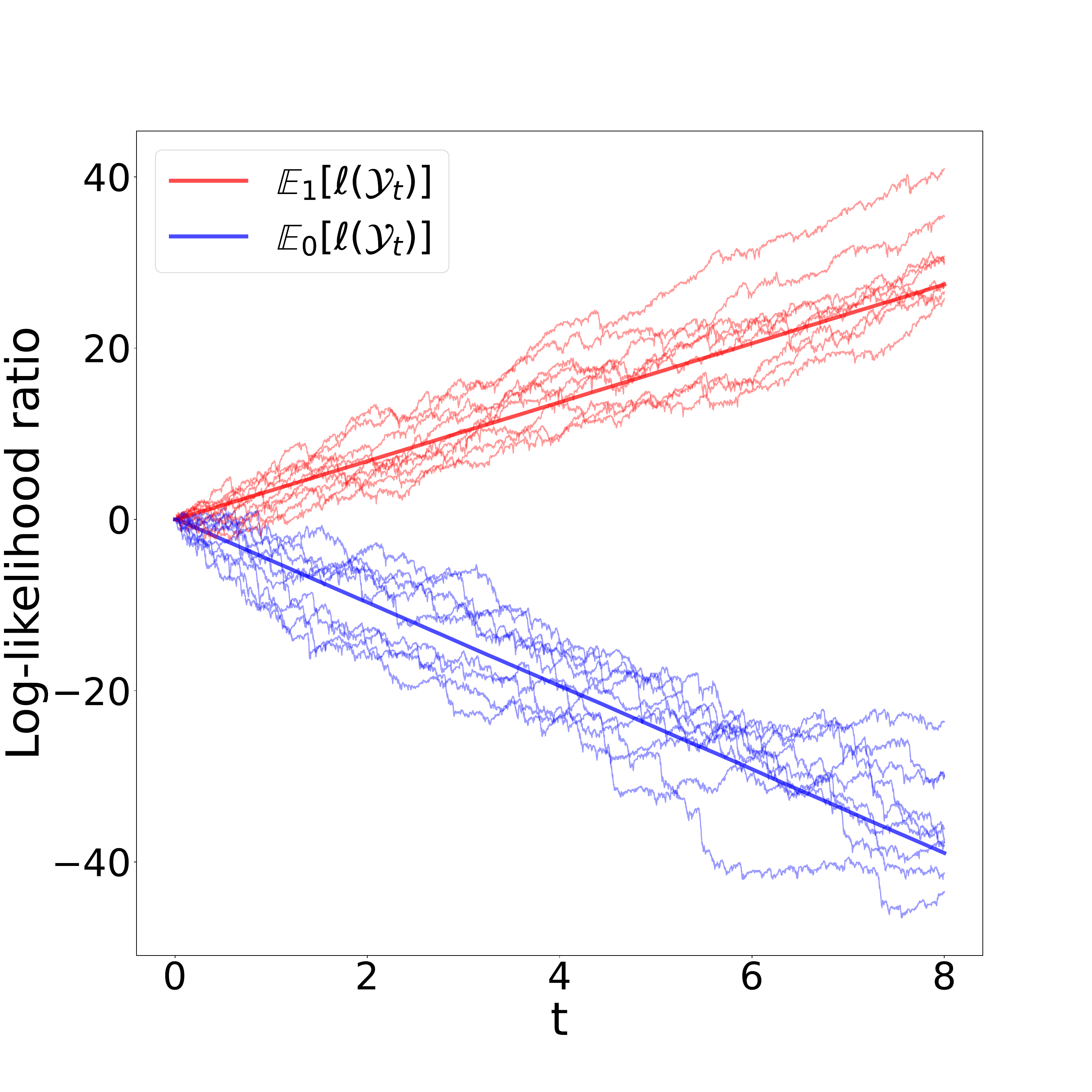}
\caption{\textit{Damping-rate discrimination}. We show the evolution of the mean-value of the LLR, obtained by sampling $\mathcal{Y}_t$ under hypothesis $h_{1(0)}$ in red(blue), along with different realizations of the stochastic process. We observe the corresponding tendency towards positive(negative) values according to the underlying true hypothesis $h_{1(0)}$.}
\label{fig:damping_realisation}
\end{figure}

As a \textit{first case study}, we make use of the SPRT to discriminate two different values of the decoherence rate $\gamma$, via a demodulated homodyne signal in the rotating frame of the mechanical frequency~\cite{Szorkovszky2011mechanical,Doherty2012quantum}.
This setting is experimentally achievable~\cite{rossi2018measurement}, and analytically treatable.

Within the rotating wave approximation, the coefficient matrices in \eqref{eq:LSE} reduce to
\begin{align}
A = -\frac{\gamma}{2} \mathbb{1}_{2},\,\,
C = -\sqrt{4\eta\kappa} \mathbb{1}_{2},\,
D= \gamma \sigma_{uc}\mathbb{1}_{2}, \,\,
\Gamma = 0, \nonumber
\end{align}
where $\sigma_{uc} = \bar{n}_{\mathrm{th}}+\frac{1}{2}+\frac{\kappa}{\gamma}$ is the covariance of the unconditional dynamics steady state, $\bar{n}_{\mathrm{th}}$ is the average number of photons in the thermal environment, $\gamma$ is a dissipative term due to the presence of an environment in thermal equilibrium interacting with the system, $\eta$ describes the measurement efficiency, and $\kappa$ is a decoherence rate induced by the measurement.

In Fig.~\ref{fig:damping_realisation} we show the evolution of the mean value of the LLR, along with several realizations of the stochastic process, for this particular system under study. For this case study, we have considered discriminating between damping rates $\gamma_1 = 440$~Hz and $\gamma_0 = 100$~Hz, and fixed the remaining parameters to be $n=1$, $\kappa=10$~Hz and $\eta=1$ for both hypotheses.

\begin{figure}[t!]
\centering
\includegraphics[width=0.5\textwidth]{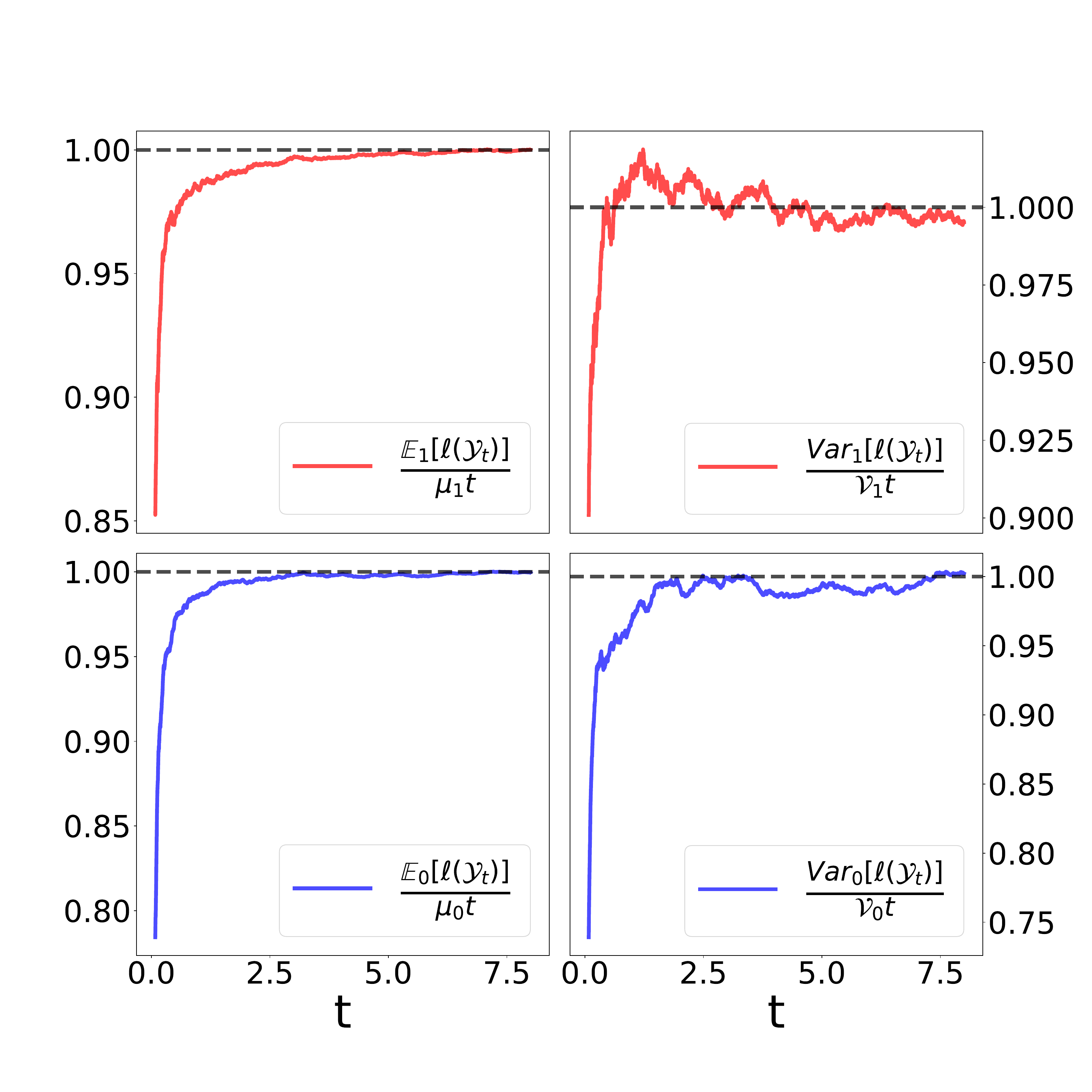}
\caption{\textit{Damping-rate discrimination}. We show the ratio between theoretical and numerically-computed values of the mean value (variance) in the first (second) column, as a function of time, when sampling from hypothesis $h_{1(0)}$ in the first (second) row; these results were computed over $2\cdot 10^{4}$ quantum trajectories. As observed, after a transient time, the first and second moments converge to the predicted analytical values.}
\label{fig:damping_moments}
\end{figure}

As discussed in Sec.~\ref{sec:hypothesis_testing}, the LLR is the main object to study in testing scenarios. As we show in the Supplemental Material, for this system we can prove that
\begin{align}\label{eq:dissmodecon}
	&\lim_{t\to \infty}\frac{\mathbb{E}_{k}[\ell(\mathcal{Y}_{t})]}{t}  =  (-)^{k\oplus 1}\mu_{k} \nonumber\\
	&\lim_{t\to \infty}\frac{\mathbb{E}_{k}[\ell(\mathcal{Y}_{t})^{2}]-\mathbb{E}_k[\ell(\mathcal{Y}_{t})]^{2}}{t} = \nu_{k}
\end{align}
where closed expressions for $\mu_{k}$ and $\nu_{k}$ can be found. This is corroborated by Fig.~\ref{fig:damping_moments}, which shows good agreement between theoretical predictions and our numerics.

\begin{figure}[t!]
    \centering
    \includegraphics[width=0.52\textwidth]{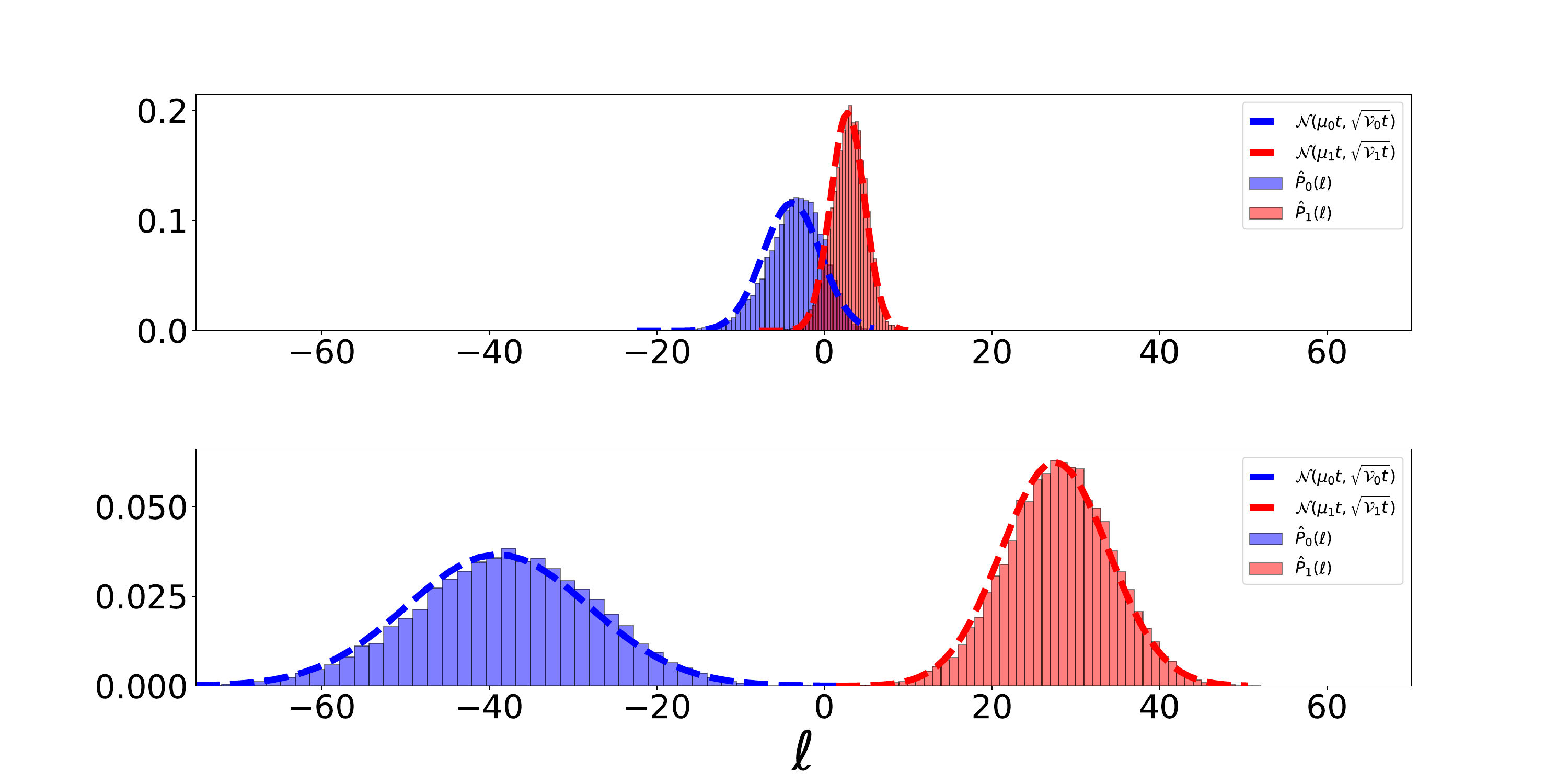}
    \caption{\textit{Damping discrimination}. We show the histogram of the LLR for time $t=2s$ ($t=8s$) in the upper (lower) panel, under hypothesis $h_{1(0)}$ in red(blue). We compare this with the theoretically predicted Gaussian distributions}.
    \label{fig:damping_distributions}
\end{figure}

With the help of Chebyshev inequality and \eqref{eq:dissmodecon}, we can readily show that conditions of theorems~\ref{th:1} and ~\ref{th:4} are satisfied, guaranteeing both weak and mean asymptotic optimality of the SPRT. This is again confirmed in Figure \ref{fig:damping_distributions} which shows how the LLR histogram under each hypothesis approaches a Gaussian distribution with the predicted drift and variance.

In order to illustrate how to implement the SPRT, and prove the advantage of sequential against deterministic tests, we have carried out a numerical experiment  \cite{githubCMON} by simulating a large number $N$ of stochastic trajectories  ---integrating \eqref{eq:LSEa}--- and used \eqref{eq:linealLogy} to keep track of the LLR for each of the measurement records as one would do in a real experiment. We consider a symmetric hypothesis testing scenario with equal priors, therefore we generate $N/2$ trajectories under $h_{0}$ and the other half under $h_{1}$.

\begin{figure}[htb]
\centering
\includegraphics[width=0.5\textwidth]{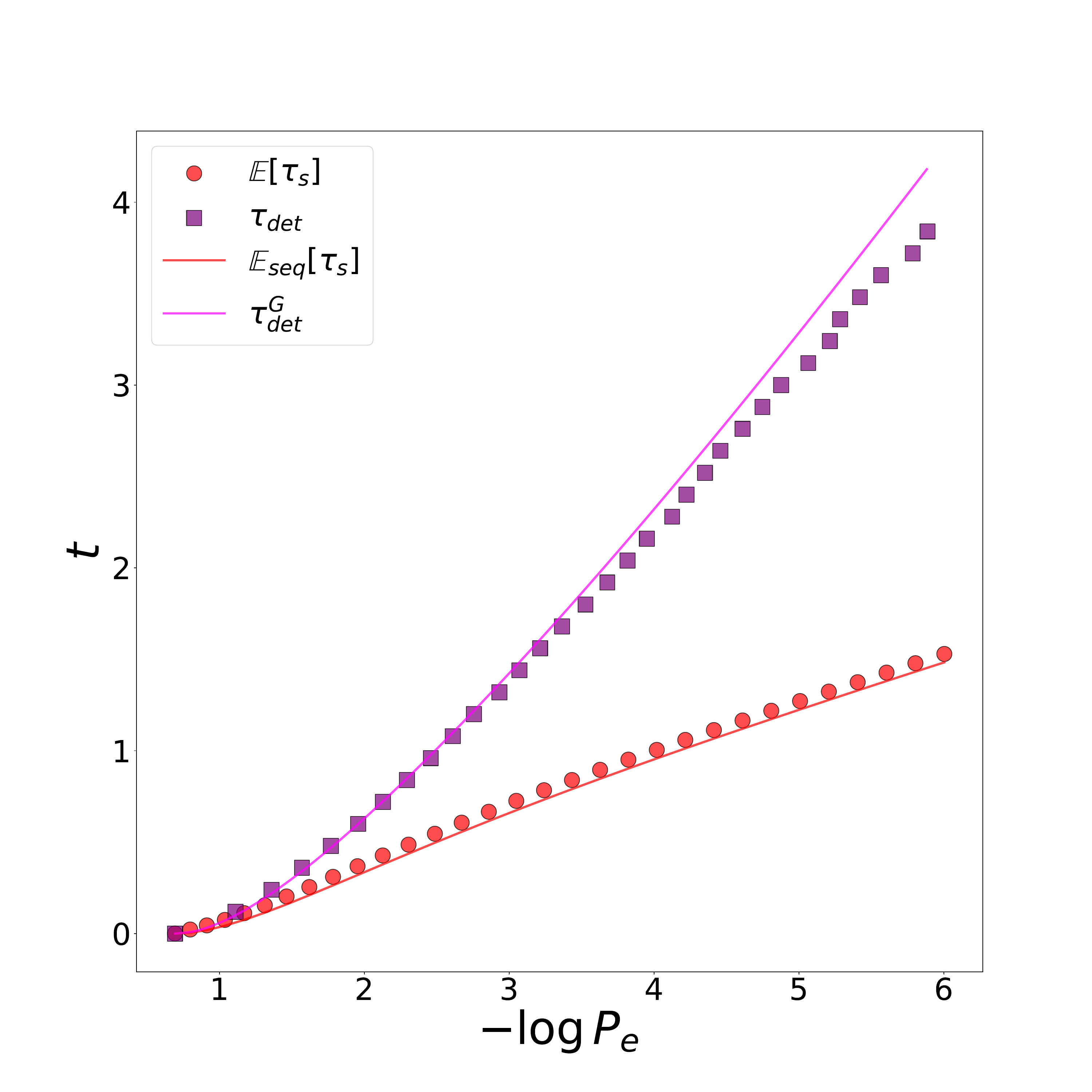}
\caption{\textit{Damping-rate discrimination}. We compare the time required for the optimal deterministic and sequential tests to reach a certain symmetric error-probability threshold. The performance is computed by averaging stopping times (sequential) and estimating the error made for each fixed time (deterministic) over $N=4\cdot 10^4$ trajectories, averaged over both hypotheses. Moreover, we compare the numerics with the theoretical predictions (solid lines) using eqs.~\eqref{eq:waldgaussian} and ~\eqref{eq:waldgaussian2} for the sequential test, and the Gaussian model for the deterministic one, here denoted by $\tau^G_{det}$, using eqs.~\eqref{eq:detgaussian1} and eqs.~\eqref{eq:detgaussian2}. The later model is clearly seen to be invalid (see main text).}
\label{fig:damping_comparison}
\end{figure}

To implement the optimal deterministic strategy,  at every time $t$ we check whether the LLR is above or below the threshold value:
 $\ell_{t}\geq a=0$ (decide in favour of $h_{1}$) or $\ell_{t}<0$ (decide in favour of $h_{1}$). We keep a record of the number of incorrect guesses $N_{F}$ and estimate
 $P_{\mathrm{err}}\approx P_{e}:=\tfrac{N_{F}}{N}$. In Fig.~\ref{fig:damping_comparison} we show the so obtained values $(-\log{P_{e}},t)$ for various times.

To implement the optimal sequential strategies, we apply the
  SPRT by fixing the equal upper and lower thresholds $a_{0}=a_{1}=\frac{1-\epsilon}{\epsilon}$ and for each trajectory $i$ we keep a record of the times $\tau_{i}$ when the LLR first hits the boundary, and of the wrong guesses (cases where the upper threshold is hit but the true hypothesis was  $h_{0}$ or vice versa). The mean stopping time is estimated as $\expect(\tau_{\mathrm{s}})\approx \bar\tau=\tfrac{1}{N}\sum_{i}\tau_{i}$, and in Fig.~\ref{fig:damping_comparison}  we plot  the points with coordinates  $(-\log \epsilon,t=\bar\tau)$  for a range of values of $\epsilon$.

\begin{figure}[htb!]
\centering
\includegraphics[width=0.5\textwidth]{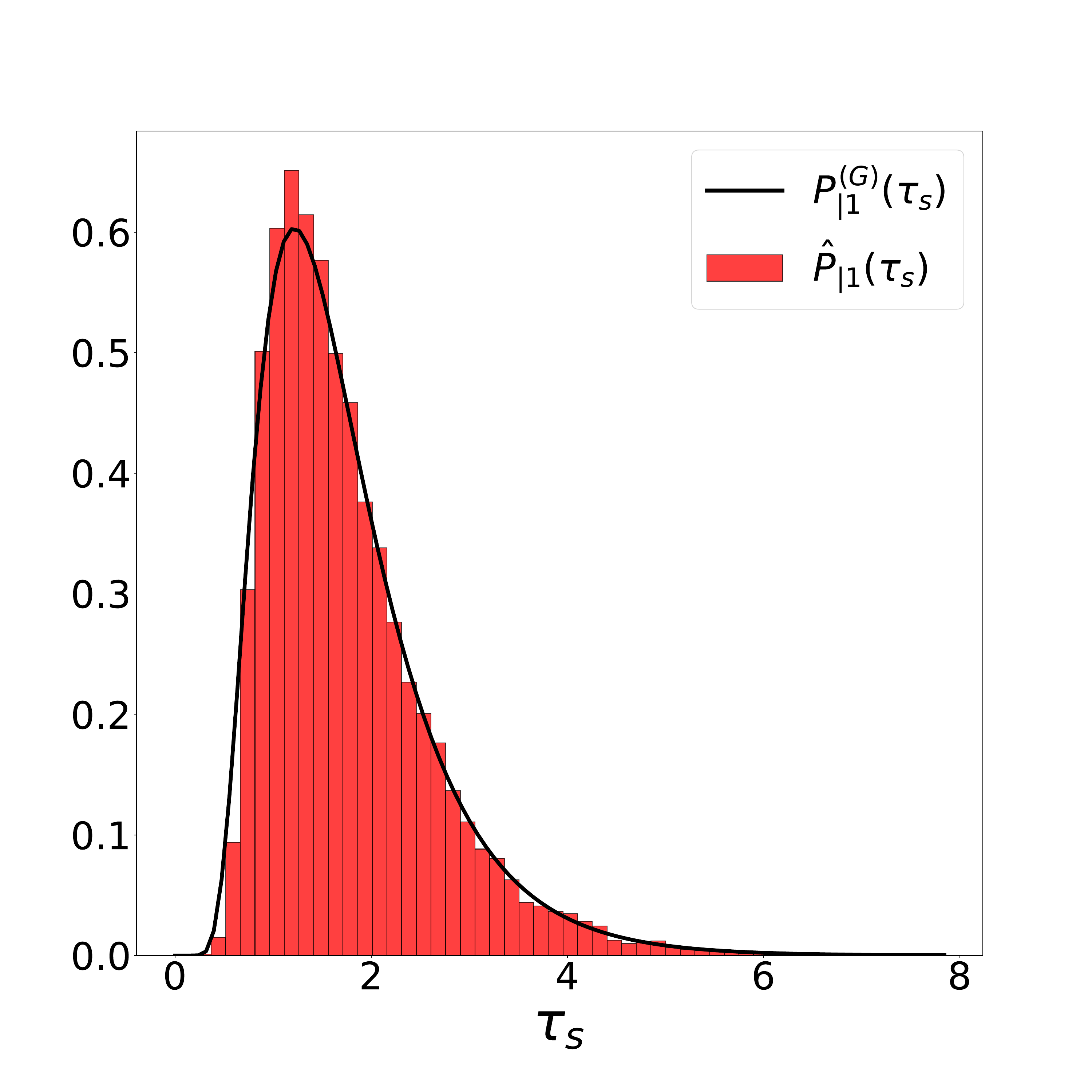}
\caption{\textit{Damping-rate discrimination}. We compute the histogram of the stopping times for the sequential test and compare it with the Inverse Gaussian probability distribution of theorem~\ref{th:43}, showing a good agreement. Results are shown when $h_1$ holds true (similar results are obtained when swapping the underlying hypothesis).}
\label{fig:Wald_distr}
\end{figure}

Before discussing the theoretical curves shown in  Fig.~\ref{fig:damping_comparison}, we can already highlight the distinct advantage of the sequential strategy over the deterministic one: the duration of the experiment required to identify the true hypothesis with a given error probability is about three times longer for the deterministic strategy than the (average) elapsed time in the sequential strategy. In addition, we also note the sequential strategy is able to certify the error probability for each single trajectory,
while the deterministic protocol only guarantees the error bound when averaging over many trajectories. An important caveat of sequential strategies is that the exact duration of the experiment is unpredictable. The mean stopping time is bounded (and may be known beforehand). However, as shown in Fig. \ref{fig:Wald_distr} the stopping time distribution has quite long tails, so a particular experiment may take substantially longer than expected. The figure also shows an excellent fit with the theoretical curve given in \eqref{eq:STdist}.

We now proceed to discuss the theoretical curves displayed in Fig.\ref{fig:damping_comparison}. Since the mean stopping time depends on the underlying hypothesis and we have assumed an equally likely hypothesis, the average stopping time is written as
\be
 \expect[{\tau_{\mathrm{s}}}]=\frac{1}{2} (\expect_{0}[{\tau_{\mathrm{s}}}]+\expect_{1}[{\tau_{\mathrm{s}}}])
 \label{eq:waldgaussian}
\ee
with
  \begin{align}
&\expect_{k}[{\tau_{\mathrm{s}}}]\approx  \frac{-a_{k\oplus 1}\alpha_{k\oplus 1}+a_{k}(1-\alpha_{k\oplus 1})}{\mu_{k}}=\nonumber\\
&= \frac{a(1-2(1+\ex{a})^{-1})}{\mu_{k}}=
 \frac{\log(\tfrac{1-\epsilon}{\epsilon})(1-2\epsilon)}{\mu_{k}},
  \label{eq:waldgaussian2}
 \end{align}
 where we have used \eqref{eq:statistial_errors} to obtain $\alpha_{k}=(1+\ex{a})^{-1}=\epsilon$. This is an asymptotic result, however, it is already an excellent approximation when required stopping times are large compared to the relaxation time of $\frac{d\expect[\ell_{t}]}{dt}$ to its stationary value $\mu_{k}$ ---see section \eqref{ssec:QMONGAUSStest}.

A naive interpretation of our previous results and the histograms shown in
Fig. \ref{fig:damping_distributions} might lead to the conclusion that
for all practical purposes, one can assume that the LLR is a Gaussian distribution with mean $\mu_{k} t$ and variance $\mathcal{\nu}_{k} t$.
Under these assumptions we can easily compute the probability of error of the deterministic strategy using the results in Sec. \ref{ssec:gaussian_iid}. Indeed, since
\be
\label{eq:detgaussian1}
P_{\mathrm{err}}=\frac{1}{2}(\alpha_{0}+\alpha_{1}),
\ee
 using first equalities in \eqref{eq:typeErrorG} with $a=0$,
\begin{align}\label{eq:detgaussian2}
	\alpha_{0}=
	\frac{1}{2} \erfc\left[\frac{\mu_{1} \sqrt{t}}{\sqrt{2 \nu_{1}}}\right];
	\alpha_{1}&=\frac{1}{2} \erfc\left[\frac{\mu_{0} \sqrt{t}}{\sqrt{2\nu_{0}} }\right].
\end{align}
 {However, this result is bluntly wrong as it is apparent from the mismatch shown in Fig. \ref{fig:damping_comparison}  between the
numerical simulation results and the (magenta) curve $(-\log P_\mathrm{err}, t)$ obtained from
\eqref{eq:detgaussian1}}.
The following simple theorem highlights that
the underlying assumption above results, namely the Gaussianity of $\ell$, the reason for this discrepancy.

\begin{theorem}\label{th:gaussian_iid}
If the log-likelihood ratio $\ell (x):=\log \frac{p_{1}(x)}{p_{0}(x)}$
is Gaussian distributed random variable under one of the hypotheses, then it will be Gaussian  distributed under both hypothesis and their means and the variances must fulfill the following relation
\begin{align}\label{eq:gauss1}
\mathbb{E}_{k}[ \ell ] &= (-)^{k\oplus 1}\mu\nonumber\\
\Var_{0}[\ell] &=\Var_{1}[\ell]=2 \mu
\end{align}
with $\mu > 0$.
i.e. PDF for $\ell$ is given by
\begin{align}\label{eq_gauss2}
p_{1}(\ell) = p_{0}(-\ell)=\frac{1}{\sqrt{4\pi \mu}} \ex{-\tfrac{(\ell- \mu)^{2}}{4\mu}}.
\end{align}
\end{theorem}
\begin{proof}
Take $p_{1}(\ell)$ to be the Gaussian probability distribution, with mean $\mu>0$ and variance $\sigma^{2}$ . From the definition of $\ell(X)$ it immediately follows
that $p_{0}(\ell)=\ex{-\ell} p_{1}(\ell)$, which means that the distribution of $p_{1}(\ell)$ has also quadratic exponent. Imposing the normalization condition on  $\ex{-\ell} p_{1}(\ell)$ we readily obtain condition $\sigma^{2}=2\mu$, and the rest of the claims follow.  More succinctly,  the following relation between the moment-generating functions holds true:
\begin{align}
    \chi_{0}(q)&:= \mathbb{E}_{0}[\ex{q\ell}]=
    \mathbb{E}_{1}[\ex{-\ell}\ex{q\ell}]=\nonumber\\
&=\mathbb{E}_{1}[\ex{(q-1)\ell}]=:\chi_{1}(q-1).
\end{align}
Recalling the moment-generating function for a Gaussian distribution to be $\chi(q)= \exp(q\mu +\frac{q^2}{2}{\sigma^{2}})$ we obtain  $-q\mu_{0}+\frac{q^{2}}{2}\mathcal{V}_{0}=(q-1)\mu_{1}+\frac{(q-1)^{2}}{2}\mathcal{V}_{1} $. Taking, e.g.,  $q=0,1$ (i.e. the normalization conditions) together with $q=1/2$ leads to desired results.
\end{proof}

It is easy to verify that these conditions do not hold in general ---in the current case under study it is sufficient to check that $\mu_{0}\neq\mu_{1}$ as shown in \eqref{eq:mukANAL} of the Supplemental Material. Therefore, it is clear that the underlying assumption that led to \eqref{eq:detgaussian2} does not hold.  It is important to emphasize that this fact does not contradict the results presented in this paper, nor does it undermine the content of the theorems stated in the previous section. This is because convergence in the probability of a specific stochastic variable does not necessarily imply convergence in moments. While the theorems address the concentration properties (i.e., the bulk) of the LLR distribution, that contribute for instance to the mean stopping time,  the calculation of the deterministic error rates involves large deviation properties of the PDF (i.e., the tails), which are not expected to follow a Gaussian distribution.

 {It is worth mentioning that the seminal paper on (deterministic) hypothesis testing in continuously monitored systems \cite{Tsang2012Continuous}, specifically in its Supplementary Material, provides integral expressions for the Chernoff Information  (see also \cite{kazakos_spectral_1980, trees_detection_2001}), which provides upper-bounds to the deterministic probability of error.}

\begin{figure}[h!]
    \centering
    \includegraphics[width=0.5\textwidth]{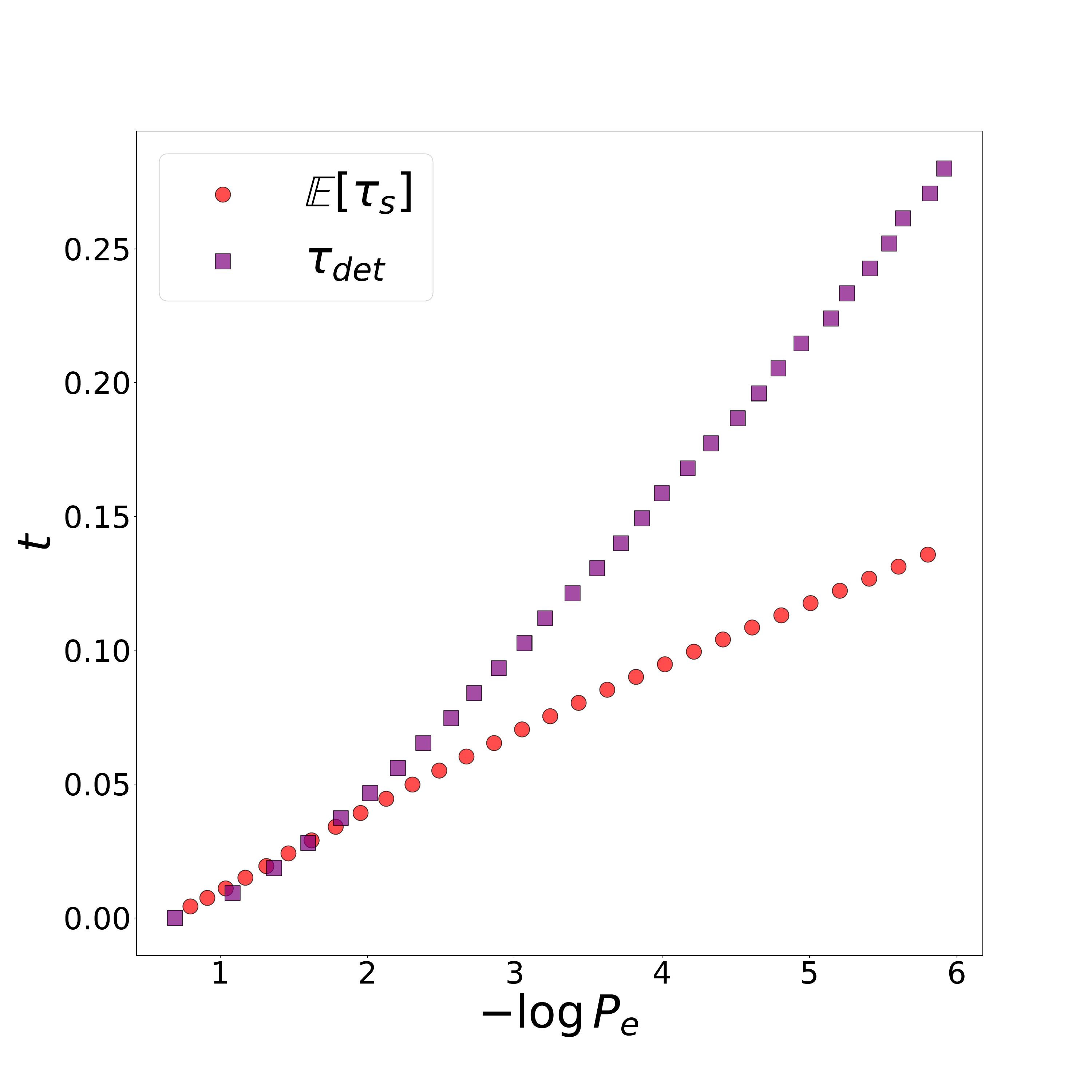}
    \caption{\textit{Frequency discrimination}. We compare the average time required for the sequential test to reach a predefined error threshold $P_\mathrm{err}$, with the time for which a deterministic strategy would need to attain such error threshold. These quantities are estimated using $2\cdot 10^4$ simulated quantum trajectories for each hypothesis, and averaging the performance over both hypotheses.}
    \label{fig:frequency_comparison}
\end{figure}

As a \textit{second case study}, we consider the discrimination between two different values of the oscillation frequency $\omega_{m}$ of the mechanical mode. Within the rotating wave approximation, the system is described by the set of differential equations in \eqref{eq:LSE}, with coefficient matrices given by

\begin{align}
A =& \Big(\;\begin{matrix}-\frac{\gamma}{2}& -\omega\\\omega & -\frac{\gamma}{2}\end{matrix}\;\Big),\,
 &&C = \sqrt{4 \eta \kappa} \Big(\begin{matrix}1&0\\0&0\end{matrix}\Big),\nonumber\\
D =& \gamma \sigma_{uc}\one_{2},\, &&\Gamma = 0,\nonumber\\
\end{align}
where $\omega$ is the mechanical-mode frequency and the remaining parameters are defined above. The frequency values have been set to $\omega_1 = 1.02\;\omega_0$ and $\omega_0 = 10^4$~Hz, while the values of the remaining parameters are $\gamma = 500$~Hz, $\kappa = 10^3$, $n=1$ and $\eta = 1$ for both hypothesis. In this case, we are not able to find an analytical expression for the variance $\sigma_{k}$ --- already the solution of Riccati equation has a transcendental solution. However, numerical results show that the variances $\sigma_{k,t}$  admit an asymptotic stationary state, similar to eqs.~\eqref{eq:dissmodecon}.

 Similarly to the damping-rate discrimination, we have carried out a comparison between deterministic and sequential tests, as shown in Fig.~\ref{fig:frequency_comparison}. A clear advantage in favor of the latter strategy is also observed.

\begin{figure}[h!]
    \centering
    \includegraphics[width=0.5 \textwidth]{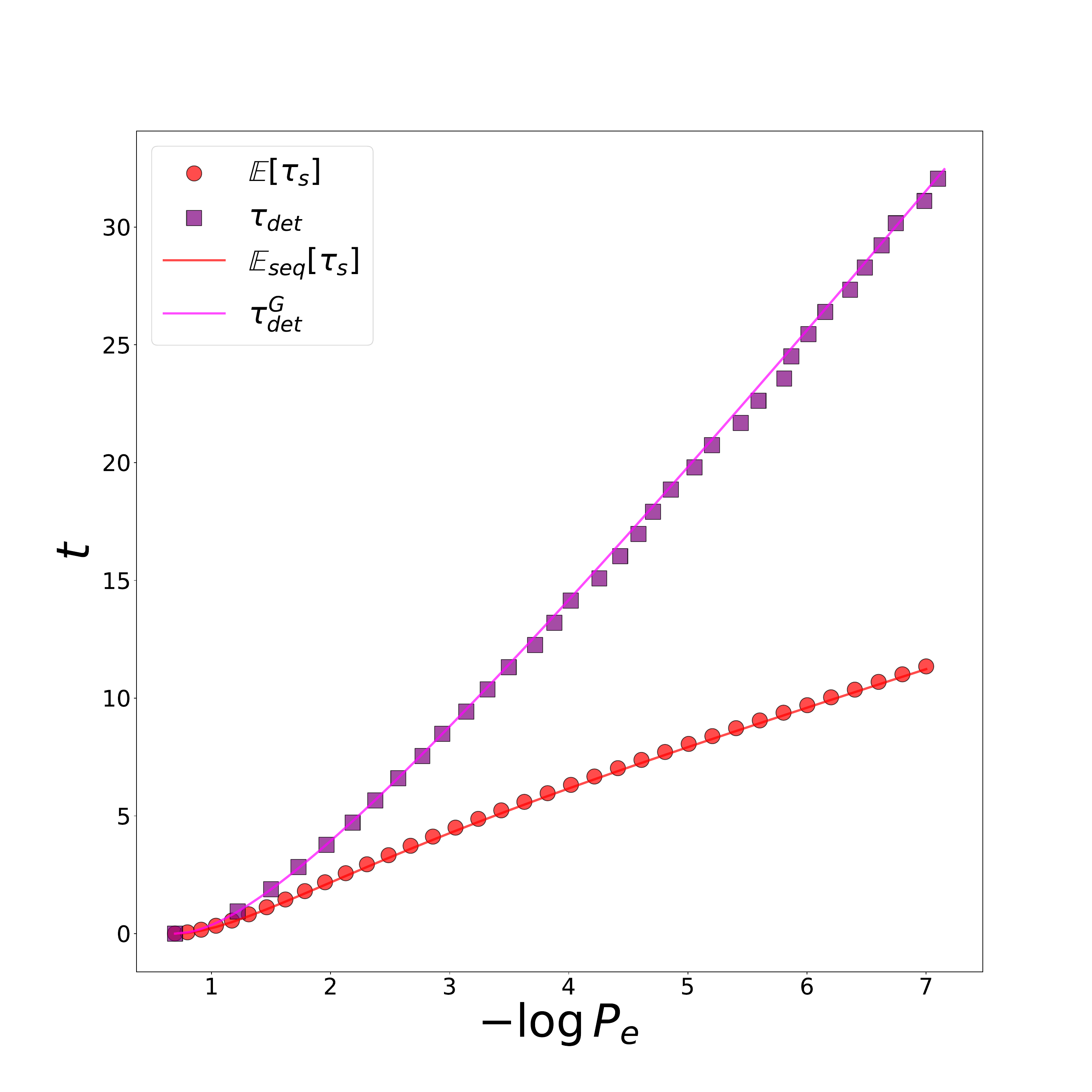}
     \caption{\textit{Force discrimination}. We compare the sequential and deterministic test, over $4\cdot 10^4$ trajectories.
     Simulated performance of both tests is compared with the theoretical predictions (solid lines) using eqs.~\eqref{eq:waldgaussian} and ~\eqref{eq:waldgaussian2} for the sequential test, and the Gaussian model for the deterministic one, here denoted by $\tau^G_{det}$, using eqs.~\eqref{eq:detgaussian1} and eqs.~\eqref{eq:detgaussian2}. The latter model is seen to be valid, in this case (see main text)}
    \label{fig:force_comparison}
\end{figure}

\textit{Force discrimination}. As a final example, we consider how well can a test do when it comes to detecting the presence of an external force. This enters as a linear term in the first-moment dynamics, as described by $\mathbf{b}_{\theta,t}=(0,b_\theta)^T$ in \eqref{eq:LSE}. Here, we have considered a constant force, with values $b_1 = 40 $Hz and $b_0 = 0$; the remaining parameters were set to $\gamma = 500$~Hz, $\kappa=10$~Hz, $\omega = 10^2$~Hz, $n=1$ and $\eta = 1$. Figure~\ref{fig:force_comparison} shows the corresponding numerical simulations in an error vs time plot, where sequential gives a similar advantage as the other cases.
{In this particular instance, hypothesis $h_1$ is indistinguishable from hypothesis $h_0$ except for a displacement in the origin of the harmonic oscillator. Consequently, we expect that the multivariate Gaussians characterizing the statistical properties of the observed signal will be identical for both hypotheses, except for their means, which will differ.
From here, it is not hard to show that, as in the example in  \ref{ssec:gaussian_iid}, the log-likelihood ratio is a linear function of the (Gaussian) signal ($\mathcal{Y}_t$) and therefore it is itself a Gaussian stochastic variable.  The Gaussianity of $\ell_{t}$ allows one to use \eqref{eq:waldgaussian2} to calculate the mean stopping time for the optimal sequential strategy and \eqref{eq:detgaussian1} to determine the probability of error for the optimal deterministic strategy. Furthermore, the Gaussianity of $\ell_{t}$ also implies that all its statistical properties under both hypotheses can be defined by a single parameter $\mu$ (as indicated in Theorem \ref{th:gaussian_iid}). The analytical curves resulting from these considerations are illustrated in Figure~\ref{fig:force_comparison}.
One can easily derive the asymptotic expression of these curves to find that in the limit of small error probability $\epsilon$, the mean stopping time for sequential strategies is four times shorter than the required time for a deterministic strategy: $T_{\det} = 4 \frac{\log\epsilon}{\mu} \sim 4\expect[\tau_{\mathrm{s}}]$.}

\section{Discussion and outlook}\label{sec:outlook}
 This study represents a primary exploration of sequential hypothesis testing in open and continuously monitored quantum systems.
 We have introduced the sequential framework and methodologies, deriving general results on statistical properties of the stopping times, a key figure of merit in sequential hypothesis testing. Moreover, we have established the optimality of the SPRT (Sequential Probability Ratio Test) for hypothesis testing under weak error conditions. Explicit closed-form expressions for the (asymptotic) mean stopping time have been provided for ubiquitous Gaussian systems, under stationarity conditions on the dynamics. Additionally, we have conducted case studies in optomechanical systems to supplement our analysis, demonstrating a clear advantage of sequential strategies over deterministic ones, with a reduction in the required measuring time by a factor between 3 and 4.

Current research efforts are focused on various extensions of this work. In particular, studying the performance of the SPRT for other detection schemes, such as performing photon-counting measurements on the leaked cavity modes, instead of dyning measurements considered in this work. A dynamical equation for the log-likelihood ratio in such scenarios has already been derived in \cite{Tsang2012Continuous}. Furthermore, we aim to determine the ultimate quantum limits for general measurement schemes, in the spirit of recent works on sequential hypothesis testing for IID quantum states \cite{QuantumVargas2021,li_optimal_2022}, or \cite{molmer_hypothesis_2015} in the context of deterministic hypothesis testing strategies for continuously monitored quantum systems.

Incorporating physically sound feedback schemes into sequential methodologies is another area of interest, as it has the potential to enhance the power of the tests. Lastly, we would like to highlight the utility of sequential analysis tools in other relevant primitives or applications, such as quickest change point detection \cite{fanizza_ultimate_2023} or anomaly detection. These tasks can be considered as genuine sequential problems, as they cannot be adequately addressed with fixed-horizon strategies.

Finally, a compelling avenue for future investigation lies in exploring model-free schemes, such as machine-learning approaches, to infer the Log-Likelihood Ratio (LLR) value without relying on a perfect model~\cite{Fallani2022learning}.

\acknowledgments
We thank Marco Fanizza for useful insights on Theorem 1.
 The numerical analysis presented in this paper was conducted using computational resources at Port d’nformaci\'o Cient\'ifica.
This work was supported by the QuantERA grant C’MON-QSENS!, by Spanish MICINN PCI2019-111869-2 and  Spanish Agencia Estatal de Investigación, project
{PID2022-141283NB-I00 with the support of FEDER funds, by the Spanish MCIN with funding from European Union NextGenerationEU (grant PRTR-C17.I1) and the Generalitat de Catalunya, as well as the Spanish MTDFP through the QUANTUM ENIA project call - Quantum Spain project, and by the European Union through the Recovery, Transformation and Resilience Plan - NextGenerationEU within the framework of the Digital Spain 2026 Agenda.
ERS was supported by FI AGAUR grant Joan Oró (ref 2023 FI-1 00410). JC also acknowledges support from the ICREA Academia award. MB also acknowledges support from AGAUR Grant no. 2023 INV-2 00034 funded by he European Union, Next Generation EU and Grant PID2021-126808OB-I00 funded by MCIN/AEI/ 10.13039/501100011033 and by ERDF A way of making Europe}

\bibliography{library}
\bibliographystyle{quantum}

\clearpage

\newpage
\onecolumn\newpage
\appendix
\setcounter{equation}{0}
\renewcommand{\theequation}{S\arabic{equation}}


\section{Asymptotic behavior of the log-likelihood ratio in the Gaussian regime}
\label{sec:agr}
The results presented in sec.~\ref{ssec:QMONGAUSStest} of the main text exploit the standard treatment for an Osrtein-Uhlenbech process to obtain, under some easy-to-verify conditions, an asymptotic expression for log-likelihood's drift and consequently of the Sequential Probability Ratio Test (SPRT) stopping time mean value.

However, having knowledge solely about the asymptotic behavior of the log-likelihood drift is insufficient to guarantee the optimality of the test in the $\mathbf{C}(\alpha_{0},\alpha_{1})$ class or to determine the probability distribution of the SPRT  stopping time in the asymptotic regime.

To achieve this goal, it is necessary to possess at least some understanding of the asymptotic behavior of the variance of the Log-Likelihood Ratio (LLR). This information is crucial in bounding the speed of convergence of the LLR by a deterministic function of time.

An expression for the mean and the variance of the LLR that can be numerically (and in some cases analytically) computes, can be obtained from the formal integral solution to the stochastic differential equation for $\mathbf{X}_{t}$, i.e.
\begin{align}
\mathbb{\Delta}^{T}\mathcal{C}\mathbf{X}_{t} &= \mathbb{\Delta}^{T}\mathcal{C}\mathcal{T}\ex{\int_{t_{0}}^{t} ds (\mathcal{A}-\chi(\Sigma_{\mathrm{s}})\Pi_{k}\mathcal{C})}\mathbf{X}_{t_{0}}
+\mathbb{\Delta}^{T}\mathcal{C}\int_{0}^{t}\mathcal{T}\ex{\int_{}^{t} ds (\mathcal{A}-\chi(\Sigma_{\mathrm{s}})\Pi_{k}\mathcal{C})}\mathbf{b}_{s}ds\nonumber\\
&\hspace{.2cm}+ \int_{t_{0}}^{t} \mathbb{\Delta}^{T}\mathcal{C} \mathcal{T}\ex{\int_{t_{0}}^{\tau} ds (\mathcal{A}-\chi(\Sigma_{\mathrm{s}})\Pi_{k}\mathcal{C})} \chi(\Sigma_{\tau})d\mathbf{w}_{\tau}.
\end{align}
Under the assumption that $\Sigma_{t}$ admits an asymptotic steady state~\footnote{Notice that the dynamical equation of $\Sigma$ is a Riccati equation that we know admits asymptotic solutions in specific regimes. See main text.} $\Sigma_{st}$ and $\Re[\mathcal{A}-\chi(\Sigma_{st})\Pi_{k}\mathcal{C}] < 0$ then the asymptotic mean and variance of the log-likelihood can be written as:
\begin{align}
\expect_{k}[\ell_{t}] &=
\lim_{t_{0} \to -\infty } \frac{(-)^{k\oplus 1}}{2}\left\{\left(\int_{t_{0}}^{t}g(\tau-t_{0})d\tau\right)^{2} -\int_{t_{0}}^{t}d\tau \int_{t_{0}}^{\tau} f(\tau-s)^{2}ds\right\} \nonumber\\
\expect_{k}[\ell_{t}^{2}]-\expect_{k}[\ell_{t}]^{2} &=
\lim_{t_{0} \to -\infty } \int_{t_{0}}^{t}d\tau\int_{t_{0}}^{\tau}f(\tau-s)^{2}ds 
-2\int_{t_{0}}^{t}d\tau_{1}\int_{t_{0}}^{\tau_{1}}d\tau_{2}f(\tau_{1}-\tau_{2})\int_{t_0}^{\tau_{2}}dsf(\tau_{1}-s)f(\tau_{2}-s)\nonumber\\
&+\int_{t_{0}}^{t}d\tau_{1}\int_{t_{0}}^{\tau_{1}}d\tau_{2}\left(\int_{t_{0}}^{\tau_{2}}f(\tau_{1}-s)f(\tau_{2}-s)ds\right)^{2},
\end{align}
where
\begin{align}
g(t-t_{0}) =    \mathbb{\Delta}^{T}\mathcal{C}\ex{(\mathcal{A}-\chi(\Sigma_{st})\Pi_{k}\mathcal{C})(t-t_{0})}\mathbf{X}_{t_{0}}, \hspace{1cm}
f(t-s) = \mathbb{\Delta}^{T}\mathcal{C}\, \ex{(\mathcal{A}-\chi(\Sigma_{st})\Pi_{k} \mathcal{C})(t-s)} \chi(\Sigma_{st})\mathbb{Q}
\end{align}
with $\mathbb{Q}= (\one,\one)^{T}$.
 As one can see from the above expression, the variance of the LLR does not, in general, fulfill the condition in ~\eqref{eq:gauss1} that any LLR with Gaussian statistics must obey. The complexity of the equations makes it challenging to derive sufficient conditions for the conditions in ~\eqref{eq:gauss1} to hold.
So we move to study the specific cases analyzed in the main text.

\subsubsection{Case study in section \ref{sec:linear_opto}}
For the first case study  it is not difficult to check the stability condition $\Re[\mathcal{A}-\chi(\Sigma_{st})\Pi_{k}\mathcal{C}] < 0$ is satisfied. Hence,  the Riccati equation describing the evolution of the covariance matrix under each of the hypotheses, admits a stationary solution, which is given by the diagonal matrix:
\begin{align}
	\sigma_{k} =  \frac{\gamma_{k}}{8\eta\kappa_{k}}\left( \sqrt{1+\frac{16\eta\kappa_{k}}{\gamma_{k}} \sigma_{uc,k}^{2}} -1\right)\mathbb{1}_{2},
\end{align}
where $\sigma_{uc,k} = \bar{n}_{th,k}+1/2+\kappa_{k}/\gamma_{k}$
is the covariance of the steady state of unconditional dynamics.
This fact guarantees that  $\Sigma_{t}$ will tend to the steady state  $\Sigma_{st}= \sigma_{0}\oplus\sigma_{1}$.
With this in hand and after some calculations one arrives to the following solutions for the mean
{
\begin{align}\label{eq:mukANAL}
\mu_1 = \frac{c^2(\gamma^2_1\chi^2_0 + 2c\gamma_0(\chi_0 - \chi_1)^2 + \gamma_0^2\chi_1^2 + \gamma_0\gamma_1(\chi_0^2 - 4\chi_0\chi_1 + \chi_1^2)}{\gamma_0(\gamma_1 + 2 c \chi_1)(\gamma_0 + \gamma_1 + 2 c \chi_1)},
\end{align}
where $\chi_k = c \sigma_k$, with $\sigma_k$} and a similar expression for the variance, can be obtained (see Ref.~\cite{githubCMON}), showing that conditions in \eqref{eq:dissmodecon} are satisfied.

\section{Stopping time probability for a 1-D bounded stochastic process}
\label{sec:Wald}

Let $Y_{t}$ be a 1-D stochastic process described by the stochastic differential equation in the It\^{o} form:
\begin{align}\label{eq:stY00}
dY_{t} = \mu_{t}(Y_{t},t)dt+\sigma(Y_{t},t)dw_{t}.
\end{align}
where $\mu(y,t)$ and $\sigma(y,t)$ are real valued continuous functions and $dw_{t}$ is the Wiener increment characterized by zero mean and covariance $\E[dw_{t}dw_{\mathrm{s}}]=\delta(t-s)dt$.

Let $\tau$ be a stopping time defined by $\tau = \inf\{t\ge 0| Y_{t} \notin (a_{0},a_{1}) \} $, i.e. the process stops when one of the two boundaries is reached.
The stopping time probability $P(t) = \E[\delta(\tau-t)]$ can be obtained as the time derivative of the stopping time cumulative distribution, i.e.
\begin{align}\label{eq:stpd0}
  P(t) = -\frac{d}{dt}P(\tau> t).
  \end{align}
The cumulative distribution gives the probability of the process to be in the interval $(a_{0},a_{1})$, i.e. $P(\tau> t) = \int_{a_{0}}^{a_{1}} \mathcal{P}(y,t)dy$ where $\mathcal{P}(y,t)\equiv  \E [\delta(Y_{\min(t,\tau)}-y)]$ is the probability distribution of the stopped process $Y_{\min(t,\tau)}$. From this one obtains:
\begin{align}\label{eq:stpd}
P( t) = -\frac{d}{dt}\left(\int_{a}^{b}dy \mathcal{P}(y,t) \right).
\end{align}
The evolution of the probability ($\mathcal{P}(y,t)$) associated with the stopped process $Y_{\min(t,\tau)}$ is described by the Fokker Planck equation associated to the stochastic process $Y_{t}$ plus absorbing boundary conditions, i.e. by:
\begin{enumerate}
\item $\partial_{t}\mathcal{P}(y,t) = \partial_{y}\mu(y,t) \mathcal{P}(y,t) +\frac{1}{2}\partial_{y}^{2} \sigma(y,t)^{2}\mathcal{P}(y,t)$, i.e. the Fokker Planck equation associated to the stochastic process described by \eqref{eq:stY00}.
\item $\mathcal{P}_{0}(y) = P_{0}(y)$, s.t. $\int_{a}^{b}dy P_{0}(y)=1$.
\item $\mathcal{P}(y\notin (a_{0}, a_{1}),t)=0 $, i.e. the trajectories that reach the boundary are removed from the ensemble.
\end{enumerate}
It is worth noticing that conditions 1 and 3 can be used together with  \eqref{eq:stpd}  to obtain the following expression for the stopping time probability distribution
\begin{align}
\label{eq:ptau}
P(t) &=  \frac{1}{2}\sigma(y,t)^{2} \partial_{y}\mathcal{P}(y,t)|_{x=a_{0}}^{a_{1}}
 +\int_{a_{0}(t)}^{a_{1}(t)}( \dot{a}_{0}(t)\partial_{a_{0}(t)}+\dot{a_{1}}(t) \partial_{a_{1}(t)})\mathcal{P}(y,t).
\end{align}
\subsection{Stopping time for a Stochastic process with deterministic drift and diffusion.}\label{sec:inversegaussianpro}
In this section, we restrict to study the case where drift and diffusion term in \eqref{eq:stY00} are continuous real deterministic functions of time, i.e. $\mu(y,t)=\mu(t)$ and $\sigma(y,t)=\sigma(t)$.
In this case, the problem of finding the stopping time probability of the stochastic process $Y_{t}$ bounded in the region $\Omega=(a_{0},b_{1})$, can be mapped to the equivalent problem of finding the stopping time probability of the stochastic process $X_t =Y_t- \int_{0}^{t} \mu_{\mathrm{s}}ds$ confined in the moving region $[a_{0}(t),a_{1}(t)]$ where $a_{i}(t)= a_{i}+\int_{0}^{t} \mu_{\mathrm{s}}ds$.
The Fokker-Planck equation associated with the process $X_{t}$ is symmetric under reflection operations of the form $x\to 2\beta-x$ with $\beta \in \mathbb{R}$, allowing for the use of the image charge method to find an explicit solution in the form of an infinite series for the survival probability $\mathcal{P}(x,t)$, i.e.
\begin{align}
\label{eq: charge_series}
\mathcal{P}(x,t) =P(x,t)
+\sum_{n=1}^{\infty}&\Big\{ P(2n(a_{1}(t)-a_{0}(t))+x,t)
- P(2{n}a_{0}(t)-2(n-1)a_{1}(t)-x,t)\nonumber\\
&+P(2n(a_{0}(t)-a_{1}(t))+x,t)
- P(2{n}a_{1}(t)-2(n-1)a_{0}(t)-x,t)\Big\},
\end{align}
where
\begin{align}\label{eq:Pheat}
 P(x,t) =\frac{1}{\sqrt{2\pi \int_{0}^{t}\sigma_{\tau}^{2}d\tau}}\ex{\frac{-x^{2} }{2 \int_{0}^{t}  \sigma_{\tau}^{2}d\tau} }
\end{align}
is the solution of the differential problem $\partial_{t}P(x,t)= \frac{1}{2}\sigma_{t}^{2}\partial_{x}^{2}P(x,t)$ with initial condition $P(x,0)=\delta(x)$.
Substituting \eqref{eq: charge_series} in \eqref{eq:ptau} and after some manipulations, one may obtain the following expression for the stopping time probability distribution
\begin{align}\label{eq:stopmutsigmat}
P(t) &= \sum_{n=0}^{\infty}(\sigma_{t}^{2}\partial_{x}+2\mu_{t})P(x,t)|_{x=a_{0}(t)-2n(a_{1}-a_{0})}^{a_{1}(t)+2n(a_{1}-a_{0})}
+\sum_{n=0}^{\infty} (\sigma_{t}^{2}\partial_{x}-2\mu_{t})P(x,t)|_{x=a_{0}(t)+2(n+1)(a_{1}-a_{0})}^{a_{1}(t)-2(n+1)(a_{0}-a_{1})}.
\end{align}
In the case where the diffusion and drift are constant functions of time, i.e. $\mu_{t}=\mu$ and $\sigma_{t}=\sigma$, the eq.~\eqref{eq:stopmutsigmat} further simplifies as
\begin{align}\label{eq:stopprobser}
P(t)
&= \sum_{n=0}^{\infty} \frac{x}{ t^{3/2} \sqrt{2\pi \sigma^{2}}} \ex{\frac{-(x-\mu t)^{2} }{2 \sigma^{2} t} }
\bigg|_{x=a_{0}-2n(a_{1}-a_{0})}^{a_{1}+2n(a_{1}-a_{0})}
+\sum_{n=0}^{\infty}\frac{x-2 \mu t }{t^{3/2}\sqrt{2\pi \sigma^{2}} }\ex{\frac{-(x-\mu t)^{2} }{2  \sigma^{2} t} }
\bigg|_{x=a_{1}+2(n+1)(a_{1}-a_{0})}^{a_{1}-2(n+1)(a_{1}-a_{0})}.
\end{align}
Let us now define $a^* = \min(a_{0},a_{1})$ and take the limit $a^* \to \infty$.
Under the assumption that $\mu>0$, the asymptotic behavior of the stopping time probability is described by the inverse Gaussian distribution (also known as Wald distribution)
\begin{align}
P(t) \sim \frac{a_{1}}{t^{3/2}\sqrt{2\pi \sigma^{2}}}\ex{-\frac{(a_{1}-\mu t)^{2}}{2\sigma^{2}t}}.
\end{align}

Assuming instead $\mu<0$, the  asymptotic behavior of the stopping time probability distribution is described by the same inverse Gaussian distribution where $a_{1}$ has been replaced with $a_{0}$, i.e.
\begin{align}
P(t) \sim \frac{a_{0}}{t^{3/2}\sqrt{2\pi \sigma^{2}}}\ex{-\frac{(a_{0}-\mu t)^{2}}{2\sigma^{2}t}},
\end{align}

\section{Theorems and Proofs}
\begin{theorem}\label{th:pre-optepsilon}
Let $P_{i}$,and $\E_{i}$ denote the probability and the expectation under the hypothesis $h_{i}$, $\mathcal{S} =(d,\tau)$ denote a generic hypothesis test where $\tau$ is a Markov stopping time, $d=d(\mathcal{X}_{0}^{\tau})$  is a terminal decision function with values in the set $\{0,1\}$, and $X_{t}$ the sample of length $t$.
Let $\mathbf{C}(\alpha_{0},\alpha_{1})=\{\mathcal{S} : P_{0}(d=1)\le \alpha_{0},P_{1}(d=0)\le \alpha_{1}\}$ and $\ell_{\tau}\equiv \log \frac{P_{1}(\mathcal{X}_{0}^{t})}{P_{0}(X_{t})}$ the LLR. The SPRT is defined by the test $\mathcal{S}_{\mathrm{s}} = (d_{\mathrm{s}},\tau_{\mathrm{s}})$, with
\begin{align}
 \tau_{\mathrm{s}} = \inf\{t\ge0 ; \ell_{t}\notin (-a_{0},a_{1})\}
\end{align}
and
\begin{align}
d_{\mathrm{s}}=\begin{cases}
1 &\textit{if}\quad \ell_{\tau_{\mathrm{s}}}\ge a_{1}\\
0 & \textit{if}\quad \ell_{\tau_{\mathrm{s}}}\le -a_{0}\\
\end{cases}
\end{align}
then
\begin{align}\label{eq:supplweaktheor}
\lim_{\alpha^{*}\to 0 }P_{k}(\tau\ge T) \ge \lim_{\alpha^{*}\to 0 } P_{k}(\tau_{\mathrm{s}}^{\epsilon}\ge T)
\end{align}
where $T\in[0,\infty)$ is a generic time, $ \alpha^{*} =\max\{\alpha_{0},\alpha_{1}\}$ and $\tau_{\mathrm{s}}^{\epsilon}$ denote the SPRT stopping time in the class  $\mathcal{S}_{\epsilon} \in  \mathbf{C}(\alpha_{0}^{1-\epsilon},\alpha_{1}^{1-\epsilon})$ with $\epsilon \in (0,1)$ and $\varepsilon = \frac{\epsilon}{1+\epsilon}$.

In addition
\begin{align}
P_{k}(\tau\ge \tau_{\mathrm{s}}^{\epsilon}) = 1-\mathcal{O}(\alpha_{1-k}^{\varepsilon}).
\end{align}
\end{theorem}
\begin{proof}
Let $\Omega = (-b_{0},b_{1})$
\begin{align}\label{eq:ineqth01}
P_{0}(d=1)&= \E_{1}[\ex{-\ell_{\tau}} \mathcal{I}_{(d=1)}] 
\ge \E_{1}[\ex{-\ell_{\tau}}\mathcal{I}_{(\tau<T)\cap(d=1)\cap(\ell_{\tau \in \Omega})} ]\nonumber\\
&\ge \ex{-b_{1}} P_{1}((\tau< T) \cap (d=1) \cap (\ell_{\tau}\in \Omega))\nonumber\\
&\ge \ex{-b_{1}}\left( P_{1}((\tau< T) \cap (d=1))-P_{1}((\tau< T)\cap (\ell_{\tau} \notin \Omega))\right)\nonumber\\
&\ge \ex{-b_{1}}\left(P_{1}(d=1) -P_{1}(\tau\ge T) -P_{1}((\tau< T)\cap (\ell_{\tau} \notin \Omega))  \right)
\end{align}
where to move from the second to the third line we have used the following set of inequalities $$P(A\cap B\cap C) = P(A\cap B)-P(A\cap B \cap  \bar{C})\ge P(A\cap B)- P(A \cap \bar{C})$$.

Exploiting the inequalities in~\eqref{eq:ineqth01}, and recalling that $\alpha_{i}\ge P_{i}(d=j)$ with $j\neq i$ one obtains the following bound for the stopping time of a generic test in the class $\mathbf{C}(\alpha_{0},\alpha_{1})$:
\begin{align}\label{eq:bop}
P_{1}(\tau \ge T) \ge 1-\alpha_{1}-\ex{b_{1}}\alpha_{0}-P_{1}((\tau< T)\cap (\ell_{\tau}\notin \Omega)).
\end{align}
What is left to do is to show that the r.h.s term can be bounded by the stopping time cumulative distribution of the SPRT.
We recall that in the SPRT:
\begin{align}\label{eq:aoa1}
a_{0} \le \log\left(\frac{1-\alpha_{0}}{\alpha_{1}}\right), \,\hspace{2cm}
a_{1} \le \log\left(\frac{1-\alpha_{1}}{\alpha_{0}}\right)\nonumber\\
\end{align}
and that the cumulative distribution of the SPRT stopping time $(\tau_{\mathrm{s}})$ is described by:
\begin{align}
P_{k}(\tau_{\mathrm{s}}\ge T) = P\left(\bigcup_{t \in [T,\infty)} \ell_{t} \in (-a_{0},a_{0})\right),\hspace{2cm}
P_{k}(\tau_{\mathrm{s}} < T) = P\left(\bigcap_{t \in (0,T]} \ell_{t} \notin (-a_{0},a_{0})\right)
\end{align}
The term $P_{1}(\tau<T \cap (\ell_{\tau}\notin\Omega))$ resembles the above expression for $P(\tau_{\mathrm{s}}\le T)$, suggesting to use the inequalities in \eqref{eq:aoa1} to characterize $b_{0}$ and $b_{1}$, however making them equal to the r.h.s. of that inequality will make the term $\ex{b_{1}}\alpha_{0}$
 converge to one in the asymptotic regime producing a trivial result.
To guarantee that $\lim_{\alpha^{*} \to\infty}\ex{b_{1}}\alpha_{0}=0$ we choose
\begin{align}\label{eq:b1b0}
&b_{0} = (1-\varepsilon)\log\left(\frac{1-\alpha_{0}}{\alpha_{1}}\right),\hspace{2cm}
&b_{1} = (1-\varepsilon)\log \left(\frac{1-\alpha_{1}}{\alpha_{0}}\right)
\end{align}
with $\varepsilon \in (0,1)$.
Under this choice for $b_{0}$ and $b_{1}$, the following inequality holds
\begin{align}
P_{1}(\tau<T \cap (\ell_{\tau}\notin\Omega))\le P_{1}(\tau_{\mathrm{s}}^{\epsilon}\le T )
\end{align}
and one obtains
\begin{align}
P_{1}(\tau\ge T )-P_{1}(\tau_{\mathrm{s}}^{\epsilon} \ge T) \ge -\alpha_{1}-\alpha_{0}^{\varepsilon}(1-\alpha_{1})^{1-\varepsilon},
\end{align}
where $\epsilon=\frac{\varepsilon}{1-\varepsilon}$.
Under the further assumption $T=\tau_{\mathrm{s}}^{\epsilon}$ the inequality simplifies as follows
\begin{align}
P_{1}(\tau\ge \tau_{\mathrm{s}}^{\epsilon} )&\ge 1-\alpha_{1}-\alpha_{0}^{\varepsilon}(1-\alpha_{1})^{1-\varepsilon}
\ge 1-\mathcal{O}(\alpha_{0}^{\varepsilon})
\end{align}
that proves the theorem for $h_{1}$ once the limit $\alpha^{*} \to 0$ is taken.
The case for $h_{0}$ is similarly proved.
\end{proof}
\begin{corollary}\label{cor:weakopt2}
If furthermore
\begin{align}\label{eq:tauconvass2}
\tau_{s} = \frac{\log(\alpha_{k})}{\mu_{k}}+\smallO_{p}(\log(\alpha_{k}))
\end{align}
then the SPRT is asymptotically optimal in a weak sense, i.e.
\begin{align}\label{eq:weakoptim2}
\lim_{\alpha^{*}\to 0}  P(\tau\ge (1-\epsilon)\tau_{s})=1
\end{align}
where $\alpha^{*}= \max(\alpha_{0},\alpha_{1})$ and  $\epsilon\in (0,1)$
\end{corollary}
\begin{proof}
Replacing $T=(1-3\epsilon)\tau_{s}$ in the inequality~\eqref{eq:supplweaktheor} one obtains
\begin{align}
\lim_{\alpha^{*}\to 0}P(\tau \ge (1-3 \epsilon )\tau_{s}) \lim_{\alpha^{*}\to 0} \ge P(\tau_{s}^{\epsilon}\ge (1-3\epsilon )\tau_{s})
\end{align}
the r.h.s. of the inequality can be lower bounded via the following chain of inequalities
\begin{align}
P(\tau_{s}^{\epsilon}\ge (1-3)\tau_{s} ) &=
 P\left(\tau_{s}^{\epsilon} -(1-2\epsilon) I_{k} \ge (1-3\epsilon)\left(\tau_{s}-\frac{1+2 \epsilon }{1-3\epsilon} I_{k}\right)\right) \nonumber\\
&\ge P(\tau_{s}^{\epsilon}-(1-\epsilon) I_{k}\ge 0) -P(\tau_{s}- (1-2\epsilon )I_{k}\ge 0)\nonumber\\
&\ge P(|\tau_{s}^{\epsilon} -I| \ge \epsilon I_{k} )-P(|\tau_{s}-(1+\epsilon)I_{k}|\ge \epsilon I_{k}).
\end{align}
Fixing $I_{k} \equiv \frac{\log(\alpha_{k})}{\mu_{k}}$, and under the assumption in~\eqref{eq:tauconvass2},  one has that $\lim_{\alpha^{*}\to 0}P(|\tau_{s}^{\epsilon} -I| \ge \epsilon I_{k} )=1$ and $\lim_{\alpha^{*}\to 0}P(|\tau_{s}-(1+\epsilon)I_{k}|\ge \epsilon I_{k}) = 0$ concluding the proof.
\end{proof}
\begin{corollary}
\label{cor:momenta_opt}
If furthermore,
\begin{align}
  \expect_{k}[\tau_{s}^{n}] = \left(\frac{\log(\alpha_{k})}{\mu_{k}}\right)^{n}
\end{align}
the SPRT is asymptotically optimal in momenta, i.e.
\begin{align}\label{eq:asymptsuppl}
\expect_{k}[\tau^{n}]\ge \expect_{k}[\tau_{\mathrm{s}}^{n}](1+\smallO(1))
\end{align}
\end{corollary}
\begin{proof}
From theorem \eqref{th:optepsilon} one has that:
\begin{align}
 \lim_{\alpha^{*}\to 0}P(\tau \ge T)= \lim_{\alpha^{*}\to 0} P(\tau_{\mathrm{s}}\ge T)
\end{align}
Replacing $T=(1-\epsilon)^2\log(\alpha_{k})$ in the above equation and exploiting \eqref{eq:tauconvass2} it is not difficult to show that
\begin{align}
P(\tau \ge (1-2\epsilon) \log(\alpha)) =1.
\end{align}
Now one only needs to use Markov inequality to obtain:
\begin{align}
  \expect_{k}[\tau^{n}] \ge (1-2n\epsilon)\left(\frac{\log(\alpha_{k})}{\mu_{k}}\right)^{n}
\end{align}
which concludes the proof.
\end{proof}

\begin{lemma}
\label{lemma1}
Let $P_{i}$ denote the probability under the hypothesis $h_{i}$, $\ell_{t}$ be a sample continuous stochastic process s.t. $\ell_{t=0}=0$.
Let $\tau$ be the stopping time defined as $ \tau_{\mathrm{s}} := \inf \{ t\ge 0| \ell_{t} \notin \Omega \} $ with $ \Omega =\{\ell_{t}=x| x\in (-a_{0},a_{1}); a_{1},a_{0} > 0\}$.
Let furthermore  $a^*= \min(a_{0},a_{1})$, and $t_{k,\pm}= \frac{a_{k}}{\mu_{k}(1 \mp \delta)}$  with $\delta \in (0,1)$ and k=\{0,1\}.
If:
\begin{align}\label{eq:ellasymptt00}
 \ell_{t} = (-)^{k\oplus 1}\mu_{k} t +\smallO_{p}(t).
\end{align}
with $\mu_{k}\in \mathbb{R}_{+}$, then
\begin{align}
\label{eq:ainfp0}
&\lim_{a^*\to \infty} P\big(\bigcap_{s\in[t_{1,+},\infty)}\ell_{s} \in \Omega\bigg) = 1, \\
&\lim_{a^*\to \infty} P\big(\bigcap_{s\in[t_{1,+},\infty)}\ell_{s} \in \Omega\bigg) = 0
\label{eq:ainfp1}
\end{align}
\end{lemma}
\begin{proof}
We only prove the case $k=1$, and analogous proof for the case $k=0$ follows.
Let us start by noting that
\begin{align}
&P_{1}(\ell_{t_{1,-}}\le a_{1}) = P_{1}(\ell_{t_{1,-}}\le \mu_{k}(1+\delta)t_{1,-}), \hspace{0,5
cm}
&P_{1}(\ell_{ t_{1,-}}\ge - a_{0}) = P_{1}(\ell_{t_{1,-}}\ge -\mu_{k}(1-\delta)t_{0,+}),
\end{align}
 condition~\eqref{eq:ellasymptt00} guarantees that the r.h.s. of both of the equations goes to one when $a^*\to \infty$, proving that
\begin{align}
\lim_{a^*\to \infty}P_{1}( \ell_{t_{1,-}}\in \Omega) =1.
\end{align}
Since the stochastic process $ \ell_{t}$ is continuous in t and $\ell_{t=0} < a_{1}$, the following is also true
\begin{align}
\lim_{a^*\to \infty} P_{1}(\ell_{s}\in \Omega) = 1 \hspace{1cm} \forall\,s \in [0,\infty)
\end{align}
proving~\eqref{eq:ainfp0}.

We now prove~\eqref{eq:ainfp1}.
\begin{align}
P\big(\bigcap_{s\in[t_{1,+},\infty)}\ell_{s} \in \Omega\bigg)&\le P(\ell_{t_{1,+}} \le  a_{1} )
= P( \ell_{t_{1,+}} \le (\mu-\delta)t_{1,+})
\end{align}
and condition~\eqref{eq:ellasymptt} guarantees that
\begin{align}
\lim_{a^*\to \infty} P(\ell_{t_{1,+}}\le \mu(1-\delta)t_{+,1}) = 0,\nonumber
\end{align}
concluding the proof.
\end{proof}

\begin{theorem}\label{sec:optminsupp}
Let $P_{i}$ be the probability under the hypothesis $h_{i}$, $\ell_{t}$, a sample continuous stochastic process under the probability measure $P_{i}$ and, such that $\ell_{t=0}=0$.
Let $\tau_{\mathrm{s}}$ be a stopping time defined as $\tau_{\mathrm{s}} := \inf\{t\ge 0| \ell_{t} \notin \Omega \} $ with $\Omega =\{\ell_{t}=x| x\in (-a_{0},a_{1}),a_{1},a_{0}>0\}$,
 if:
\begin{align}\label{eq:ellasymptt}
\ell_{t} = (-)^{k\oplus 1}\mu_{k}t +\smallO_{p_{k}}(t).
\end{align}
Then
\begin{align}\label{eq:the10}
\tau_{\mathrm{s}} = \frac{a_{k}}{\mu_{k}}+{o}_{p_{k}}(a_{k}).
\end{align}
\end{theorem}
\begin{proof}
Let us define $t_{k,\pm}= \frac{a_{k}}{\mu_{k}(1\mp\delta)}$ with $\delta\in(0,1)$, then
\begin{align}
P_{k}(|\tau_{\mathrm{s}}\frac{\mu_{k}}{a_{k}}  -1| \ge \delta)  &\le P(\tau_{\mathrm{s}} \le t_{k,-}) + P(\tau_{\mathrm{s}} \ge t_{k,+}) \nonumber\\
\end{align}
The definition of the stopping time $\tau_{\mathrm{s}}$ allows for the following identities:

\begin{align}
P_{k}(\tau_{\mathrm{s}} \ge t_{k,+})    &= P(\forall s < t_{k,+},\, \ell_{\mathrm{s}} \in \Omega)\nonumber\\
P_{k}(\tau_{\mathrm{s}} \le t_{k,-})    &= 1- P_{k}(\tau_{\mathrm{s}} > t_{k,-}) =1 - P_{k}(\forall s \le t_{k,-},\, \ell_{\mathrm{s}} \in \Omega).\nonumber
\end{align}
 Lemma \ref{lemma1} shows that the r.h.s. of the above equation tends to zero when $a^*\to\infty$ with $a^*= \min(a_{0},a_{1})$
proving that
\begin{align}\label{eq:taulim}
\lim_{a^* \to \infty}P_{k}(|\tau\frac{\mu_{k}}{a_{k}}  -1| \ge \delta) = 0
\end{align}
concluding the proof of \eqref{eq:the10}.
\end{proof}

\begin{corollary}
If $\ell_{t}$ describes the LLR then $\tau_{s}$ represents the SPRT stopping time, and the test is weakly asymptotically optimal.
\end{corollary}
\begin{proof}
one only need to notice that for the SPRT $a_{k}\sim \log(\alpha_{k})$, then corollary~\eqref{cor:weakopt} guarantees the optimality.
\end{proof}
\begin{corollary}
If furthermore
\begin{align}\label{eq:avgell}
\expect_{k}[\ell_{t}] = (-)^{k\oplus 1}\mu_{k}t +\smallO(t)
\end{align}
then the mean stopping time for the SPRT has the following asymptotic value
\begin{align}
\expect_{k}[\tau_{\mathrm{s}}] = \frac{a_{k}}{\mu_{k}} +\smallO(a_{k})
\end{align}
and the SPRT is asymptotically optimal in the mean,i.e.
\begin{align}
\expect_{k}[\tau] \ge \expect_{k}[\tau_{\mathrm{s}}]+\smallO(\log \alpha_{k})
\end{align}
\end{corollary}
\begin{proof}
Let us write $\ell_{t}$ in its integral form and make use of the \eqref{eq:avgell} to obtain the following identity set of
\begin{align}
{\expect_{k}[\ell_{\tau}]} &= \expect_{k}\left[\int _{0}^{\infty} \frac{d\ell_{t}}{dt} \one_{(\tau_{\mathrm{s} \ge t})} dt \right] 
  = \int_{0}^{\infty} \expect_{k}\left[\frac{d \ell_{s}}{ds}\right] P(\tau_{\mathrm{s}}\ge s ) ds 
  = (-)^{k\oplus 1} \mu_{k}\expect_{k}[\tau_{\mathrm{s}}]+ \zeta
\end{align}
with
\begin{align}
\zeta = \frac{1}{\mu_{k}}  \int_{0}^{\infty} \left(\expect_{k} \left[ \frac{d\ell_{t}}{dt} \right]-
 (-)^{k\oplus 1}\mu_{k}\right)P(\tau_{\mathrm{s}}\ge t ) dt .
\end{align}
From the above expression, one obtains:
\begin{align}\label{eq:stopping}
\expect_{k}[ \tau_{\mathrm{s}} ] &=  \frac{\expect_{k}[\ell_{t}]}{\mu_{k}}+ \zeta.
\end{align}
what is left to do to prove is to show that $\frac{\zeta}{a_{k}}$ goes to zero in the limit of $a^*\to \infty$.
Let us study $\frac{\zeta}{a_{k}}$ and rewrite the integral of $dt$  as the sum of the following integrals
\begin{align}
\lim_{a^*\to \infty} \frac{\zeta}{a_{k}} &= \lim_{a^*\to \infty} \int_{0}^{t_{-,k}}\left(\frac{\expect_{k} \left[ \frac{d\ell_{t}}{dt} \right]-
 (-)^{k\oplus 1}\mu_{k}}{a_{k}}\right)P(\tau_{\mathrm{s}}\ge t ) dt +\int_{t_{-,k}}^{t_{+,k}}\left(\expect_{k} \left[ \frac{d\ell_{t}}{dt} \right]-
 (-)^{k\oplus 1}\mu_{k}\right)P(\tau_{\mathrm{s}}\ge t ) dt\nonumber\\
 &+\int_{t_{+,k}}^{\infty} \left(\expect_{k} \left[ \frac{d\ell_{t}}{dt} \right]-
 (-)^{k\oplus 1}\mu_{k}\right)P(\tau_{\mathrm{s}}\ge t ) dt
\end{align}
where $t_{\pm,k}= \frac{a_{k}}{\mu_{k}(1 \mp \delta)}$ with $\delta \in (0,1)$.
Since $P(\tau_{\mathrm{s}})$ is a bounded positive function one is allowed to replace $P(\tau_{\mathrm{s}}\ge t)$ with its limit on $a^*\to \infty$ in the integral, and making use of lemma~\ref{lemma1} obtain
\begin{align}\label{eq:zeta}
\lim_{a^*\to \infty} \left|\frac{\zeta}{a_{k}} \right| &= \lim_{a^*\to \infty} |\int_{t_{-,k}}^{t_{+,k}}\left(\expect_{k} \left[ \frac{d\ell_{t}}{dt} \right](-)^{k\oplus 1}\mu_{k}\right) P(\tau_{\mathrm{s}}\ge t ) \le C(t_{k,+}-t_{k,-})| = \frac{\delta}{\mu_{k}}
\end{align}
where for the last inequality we have used the fact that the averaged increment of the log-likelihood is a bounded function, i.e. $\expect_{k}[d\ell_{t} /dt] <C$ with $C< \infty $.

Then with the help of \eqref{eq:zeta}, it is not difficult to show that
\begin{align}
\lim_{a^*\to \infty} \left|\frac{\expect_{k}[\tau]}{a_{k}}-\frac{1}{\mu_{k}}\right| \le \delta
\end{align}
concluding the proof.
\end{proof}

\begin{theorem}
Let $P_{i}$ the probability distribution under the hypothesis $h_{i} $, $\ell_{t}$, the LLR, to be described by a continuous stochastic function under the probability measure $P_{i}$. Let $\tau_{\mathrm{s}}$ be a stopping time defined as $\tau(\mathcal{Y}_{t}) = \inf\{t\ge 0| \ell_{t}(\mathcal{Y}_{t}) \notin \Omega \} $ with $\Omega= (-a_{0},a_{1})$,i.e. the SPRT stopping time.\\
If
\begin{align}\label{eq:lqse001}
\ell_{t} = \tilde{\ell}_{t}+ \smallO_{p_{k}}(\sqrt{t})
\end{align}
where $\tilde{\ell}_{t} \equiv (-)^{k\oplus 1} \mu_{k} t +\sigma_{k}\zeta_{t}$, $\mu_{k}\in \mathbb{R}^{+}$ and $\zeta_{t}$ a standard Wiener process, i.e. $\mathbb{E}[\zeta_{t}]=0$ and  $\mathbb{E}[\zeta_{t}\zeta_{\tau}]=\min(t,\tau)$,
then the probability distribution of the rescaled stopping time $\tilde{\tau}_{\textrm{s}}:=\frac{\tau_{\textrm{s}}}{a_{k}}$ is asymptotically approximated by the Inverse Gaussian distribution:
\begin{align}\label{eq:pinvsupp}
P_{k}({\tilde\tau}_{\textrm{s}}=t) \sim \frac{1}{{t}^{3/2}\sqrt{2\pi \sigma_{k}^{2}a_{k}}}\ex{-\frac{(1-\mu_{k} {t})^{2}}{2\sigma_{k}^{2}a_{k}{t}}}.
\end{align}
and the SPRT is asymptotically optimal in momenta, i.e.
\begin{align}
\expect_{k}[\tau^{n}] \ge \expect_{k}[\tau_{\mathrm{s}}^{n}](1+o(1))
\end{align}
where $\tau$ represents the stopping time of a generic test in the class $\mathbb{C}(\alpha_{0},\alpha_{1})$.
\end{theorem}

\begin{proof}

We first prove eq.~\eqref{eq:pinvsupp}.
Let $a^* = \min(a_{0},a_{1})$,
lemma \ref{lemma1} shows that $\lim_{a^* \to \infty} P_{k}(\ell_{t} \in \Omega| t \le t_{k,-} ) = 1$, from which immediately follows that for $t<t_{k,-}$
\begin{align}
&\lim_{a^*\to 0} P(|\ell_{\min(t,\tau_{\mathrm{s}})}-\ell_{t}|\ge \epsilon) = 0 \hspace{0,2cm}\forall t<t_{k,-}  \nonumber\\
&\ell_{\min(t,\tau)} \xrightarrow[a^* \to \infty ]{P} \ell_{t} \hspace{0,2cm}\forall t<t_{k,-},
\end{align}
Let now $g(a^* )$ be a continuous positive increasing function s.t. $\lim_{a^*\to \infty}g(a^*)=\infty$ and $g(a^*)< t_{k,-}$, from the hypothesis in \eqref{eq:lqse001} one has that
\begin{align}
\ell_{\min(t,\tau)}\xrightarrow[a^*\to\infty]{P}\ell_{t}\xrightarrow[a^*\to\infty]{P}\tilde{\ell}_{t}\hspace{0,5cm} \forall g(a^*)\le t\le t_{k,-}.
\end{align}
This fact together with the existence and unicity of the solution of the Fokker-Plak differential equation with absorbing boundaries $\Omega$ and initial condition $\ell_{0}=0$, allows the following identification
\begin{align}
\ell_{\min(t,\tau)}=\tilde{\ell}_{\min(t,\tau)}+\smallO_{p_{k}}(\sqrt{t})
\end{align}
Since the stopping time probability distribution is described by \eqref{eq:stpd} and $P_{k}(a_{i}=0)$,  Portmanteau lemma guarantees that
\begin{align}
\lim_{a^*\to \infty} P_{k}(\tilde\tau_{\textrm{s}}) = \tilde{P}(\tilde\tau_{\textrm{s}})
\end{align}
where $\tilde{P}(\tilde\tau_{\textrm{s}})$ is the normalized stopping time probability distribution associated to $\tilde{\ell}_{t}$ and reduces to the inverse Gaussian distribution in eq.~\eqref{eq:pinvsupp} once the limit of $a^*\to\infty$ is taken (see section~\eqref{sec:inversegaussianpro} for further details), concluding the proof.

Weak asymptotic optimality of the SPRT is guaranteed by theorem \ref{sec:optminsupp}, to prove optimality in momenta we notice the probability distribution~\eqref{eq:pinvsupp}  allows for the following asymptotic result
\begin{align}
\expect_{k}[\tau_{s}^{n}] = \left(\frac{\log a_{k}}{\mu_{k}}\right)^n(1+\smallO(1)).
\end{align}
and that $\log(a_{k})\sim \log(\alpha_{k})$ for $a^{*}\to \infty$, that is a sufficient condition of asymptotic optimality as stated in the corollary\eqref{cor:momenta_opt}.
\end{proof}

\begin{theorem}[Optimal Asymmetric weak error.]\label{sec:optima_asym}
Let $P_{i}$ the probability distribution under the hypothesis $h_{i}$, $\ell(\mathcal{Y}_{t})$ the LLR and  $\alpha_{0}= P_{0}(\ell_{t}\ge a) $, $\alpha_{1}=P_{1}(\ell_{t}\le a) $ the type I and II error associated to the deterministic log-likelihood ratio test (also known as Neyman-Pearson test), let furthermore
\begin{align}
\alpha_{k}(t)^{*} := \min\{\alpha_{k}(t) : \alpha_{k\oplus 1}(t) \le \epsilon \}
\end{align}
with $\epsilon \in [0,1)$. If
\begin{align}\label{eq:condint}
\ell_{t}(\mathcal{Y}_{t}) = (-)^{k\oplus 1}\mu_{k} t +\smallO_{p_{k}}(t)
\end{align}
with $\mu_{k}\in \mathbb{R}^{+}$ then
\begin{align}
- \frac{\log\alpha_{k}^{*}(t)}{t} = \mu_{k\oplus 1}(1+o(1)).
\end{align}
where $\mu_{k}$ can be understood as the regularized Kullback-Leibler information divergence, i.e.
\begin{align}
I(P_{k}|P_{k-1}) = \lim_{t \to \infty} \frac{1}{t} \expect_{k}\left[\log \frac{P_{k}(\mathcal{Y}_{t})}{P_{k\oplus 1}(\mathcal{Y}_{t})}\right]
\end{align}
and also represents the minimum achievable error for a \textit{deterministic test} in the \textit{asymmetric case} scenario.
\end{theorem}

\begin{proof}
We will prove it only for k=0, analogous proof follows for k=0.
From condition \eqref{eq:condint} one has that
\begin{align}\label{eq:condles}
&\lim_{t \to \infty}P_{1}((\ell_{t} \le (\mu-\delta)t) =0\\
&\lim_{t \to \infty}P_{1}((\ell_{t} < (\mu+\delta)t) =1\\
&\lim_{t \to \infty}P_{1}((\ell_{t} \ge a) \cap (t\ge a/(\mu_{1}+\delta)))  = 1\\
&\lim_{t \to \infty}P_{1}((\ell_{t} \ge a) \cap (t \le a/(\mu_{1} - \delta)))  = 0.
\end{align}
The above conditions forces, $a = (\mu-\delta)t+\smallO(t)$ to guarantee that $\alpha_{0}(t)= P_{1}(\ell_{t} \ge a(t))\le \epsilon$.

Let us rewrite the error probability $\alpha_{1}$ as follows,
\begin{align}
\alpha_{1} &= P_{0}(\ell_{t} \ge a) 
= P_{0}(a\le \ell_{t} \le a+2 \delta t)+ P_{0}(\ell_{t} \ge a+ 2 \delta t ) 
= \expect_{1}[\ex{-\ell_{t}}\mathbb{1}_{(a\le \ell_{t} \le a+2\delta t} )]+P_{0}(\ell_{t} \ge a+2 \delta t )
\end{align}
where  we used the identity $\expect_{0}[\one_{\ell_{t}\in \Omega}]=\expect_{1}[\ex{-\ell_{t}}\one_{\ell_{t}\in \Omega}]$, where $\one_{ \ell_{t}\in \Omega}$ denotes the indicator function of having $\ell_{t}$ in the set $\Omega$. \\
With the above chain of identities is not difficult to obtain the following
	\begin{align}
	\ex{-(a+2\delta t) }P_{1}(a\le \ell_{t} \le a+2 \delta t )  \le \alpha_{0}(t)\le \ex{-a}\left[ (P_{1}(a\le \ell_{t} \le a+2 \delta t )+ (P_{1}(\ell_{t}\ge a+2\delta t) \right],
	\end{align}
	and thanks to the monotonicity of the logarithm get:
	\begin{align}
	\frac{a}{t}+\log P_{1}(\ell_{t}\ge a+2\delta t) \le- \frac{1}{t} \log\left(\frac{\alpha_{0}}{P_{1}(a\le \ell_{t} \le a+2 \delta t )}\right)  \le \frac{(a+2 \delta t)}{t}.
	\end{align}
Dividing now the above expression by $t$ and assuming $a = (\mu_{1}-\delta)t$ one gets
\begin{align}\label{eq:ineq_log_stein}
\left| \frac{1}{t} \log\left(\frac{\alpha_{0}}{P_{1}(|\ell_{t}-\mu_{1}| \le \delta t )} \right)+ \mu_{1}\right| \le \delta -\log P_{1}(\ell_{t}\ge (\mu_{1}+\delta) t)
\end{align}
Since hypothesis \eqref{eq:condint} guarantees that
\begin{align}
\lim_{t\to \infty} P_{0}(\ell_{t}\ge \mu_{1}+\delta t) = 0 \nonumber\\
\lim_{t\to \infty} P_{1}(|\ell_{t}-\mu_{1}| \le \delta t) = 1
\end{align}
The limit of \eqref{eq:ineq_log_stein} reduces to:
\begin{align}
\lim_{t\to \infty}\left| \frac{1}{t}\log(\alpha^{*}_{0}(t))+\mu_{1} \right|\le \delta
\end{align}
that is the thesis.
Since the Neyman-Pearson lemma guarantees the log-likelihood test to be the deterministic test with the smallest weak error, one can also conclude that $\mu_{k}$ is a lower bound for the deterministic case scenario.

\end{proof}

\end{document}